\documentclass[12pt]{article}
\usepackage[dvipdfmx]{graphicx}
\usepackage{color,amssymb,amsmath}
\usepackage{slashed}
\usepackage[a4paper, total={16.5cm,22cm}]{geometry}

\begin{document}

\newcommand{\lrf}[2]{ \left(\frac{#1}{#2}\right)}
\newcommand{\lrfp}[3]{ \left(\frac{#1}{#2} \right)^{#3}}
\newcommand{\vev}[1]{\left\langle #1\right\rangle}

\newcommand{\TeV}{\text{TeV}}
\newcommand{\GeV}{\text{GeV}}
\newcommand{\MeV}{\text{MeV}}
\newcommand{\keV}{\text{keV}}
\newcommand{\eV}{\text{eV}}

\newcommand{\OiB}[1]{(O_\chi)_{#1,\widetilde{B}}}
\newcommand{\OiHd}[1]{(O_\chi)_{#1,\widetilde{H}_d}}
\newcommand{\OiHu}[1]{(O_\chi)_{#1,\widetilde{H}_u}}

\begin{titlepage}
\begin{flushright}
UT-15-35\\
IPMU-15-0183
\end{flushright}
\vskip 3cm
\begin{center}
{\Large \bf
Prospects for Higgs- and $Z$-resonant\\
\vspace{0.5cm}
Neutralino Dark Matter
}
\vskip 1.5cm
{
Koichi Hamaguchi$^{(a,b)}$ and 
Kazuya Ishikawa$^{(a)}$
}
\vskip 0.9cm
{\it $^{(a)}$ Department of Physics, University of Tokyo, Bunkyo-ku, Tokyo 113--0033, Japan \vspace{0.2cm}
\par
$^{(b)}$ Kavli Institute for the Physics and Mathematics of the Universe (Kavli IPMU), \\
University of Tokyo, Kashiwa 277--8583, Japan
}
\vskip 2cm
\abstract{
In the minimal supersymmetric standard models, neutralino dark matter with mass of $m_{\chi}\sim m_Z/2\sim 45~\GeV$ and $m_{\chi}\sim m_h/2\sim 62~\GeV$ can have the thermal relic abundance $\Omega_{\chi_1}h^2\simeq 0.120$ via the $Z$- and Higgs-resonant annihilations, respectively, while avoiding all the current constraints.
Phenomenology of such scenarios is determined only by three parameters, Bino mass $M_1$, Higgsino mass $\mu$, and $\tan\beta$, in the limit that all other supersymmetric particles and heavy Higgs bosons are decoupled.
In this paper, we comprehensively study the constraints and future prospects of the search for such Higgs- and $Z$-resonant neutralino dark matter.
It is shown that almost all the parameter space of the scenario will be probed complementarily by the LHC search for the chargino and neutralinos, the direct detection experiments, and the Higgs invisible decay search at the ILC.
}

\end{center}
\end{titlepage}

\section{Introduction}
\label{sec:intro}

The identity of the dark matter (DM) is one of the biggest mysteries in particle physics, astrophysics, and cosmology.
Among various DM candidates, the lightest neutralino in the supersymmetric (SUSY) extension of the Standard Model (SM) is particularly attractive, since it can have the desired thermal relic abundance with a weak scale mass.

The discovery of the 125 GeV Higgs boson~\cite{Aad:2012tfa,Chatrchyan:2012ufa} and no signal of physics beyond the SM so far at the LHC run I might imply that the SUSY particles, in particular the scalar partners of SM fermions (sfermions), are much heavier than ${\cal O}(0.1$--$1)~\TeV$.
After the Higgs discovery, therefore, such heavy sfermion scenarios have attracted attentions (see, e.g., Refs.~\cite{heavy_sfermion}).

If all sfermions are heavy, the lightest neutralino can have the correct thermal relic abundance, $\Omega_\chi\simeq 0.120$~\cite{Agashe:2014kda,1502.01589}, only in limited cases.
For instance, the pure Wino DM with a mass of $\simeq 3$ TeV can explain the DM density~\cite{hep-ph/0610249,0706.4071}, which may be probed by indirect detections~\cite{Wino-indirect}.
The (almost) pure Higgsino with a mass of $\simeq 1$ TeV is also an attractive candidate~\cite{0706.4071,1410.4549}.
In addition, the coannihilations among gauginos~\cite{gaugino-coannihilation-early,hep-ph/0601041} and the well-tempered Bino-Higgsino mixng~\cite{hep-ph/0601041,hep-ph/0004043,hep-ph/0406144} can lead to the desired DM density, with various phenomenological implications (see, e.g., Refs.~\cite{gaugino-coannihilation-recent,1211.4873} and references therein).

In this paper, we study another viable corner of the neutralino parameter space in the heavy sfermion scenario, i.e., the Higgs- and $Z$-reonant neutralino DM. 
When the mass of the lightest neutralino is close to the Higgs- or $Z$-resonance, $m_{\chi_1}\sim m_h/2$ or $m_Z/2$, it can have the correct thermal relic abundance while avoiding the constraints from the direct detection and other experiments.
Aspects of such Higgs- and $Z$-resonant neutralino DM have been investigated e.g., in Refs.~\cite{Drees:1992am,hep-ph/0406144,HZpoleDM,1410.5730}.\footnote{
Higgs- and $Z$-resonant DMs are also realized in non-SUSY models with a singlet Majorana fermion and a SU(2)-doublet Dirac fermion. See e.g.,~Refs.~\cite{non-SUSY-HZpole}.
}

We revisit this Higgs- and $Z$-reonant neutralino DM scenario and extend previous studies by comprehensively investigating the current constraints and future prospects.
We include the following constraints and prospects.
\begin{itemize}
\item relic abundance $\Omega_\chi h^2 = 0.120$~\cite{Agashe:2014kda,1502.01589}.
\item DM direct detection.
\begin{itemize}
\item constraints on the spin-independent (SI) scattering cross section from the LUX~\cite{1310.8214},
and on the spin-dependent (SD) scattering cross section from the XENON 100~\cite{1301.6620}.
\item prospects of the XENON 1T for the SI~\cite{1206.6288} and SD~\cite{1211.4573} scattering cross sections.
\end{itemize}
\item Higgs invisible decay.
\begin{itemize}
\item current constraint from global fit~\cite{1303.3570,1306.2941}.
\item expected sensitivity of the HL-LHC~\cite{1309.7925,1310.8361}, and of the ILC~\cite{1310.0763}.
\end{itemize}
\item chargino/neutralino search at the LHC.
\begin{itemize}
\item expected sensitivity at 14 TeV for 300 fb$^{-1}$ and 3000 fb$^{-1}$~\cite{ATL-PHYS-PUB-2014-010}.
\end{itemize}
\end{itemize}
Other constraints are also briefly discussed.
Constraints from the LHC run I~\cite{1402.7029} is discussed in Appendix~\ref{app:8TeV}.
We use a simplified model with only three parameters, the Bino mass $M_1$, the Higgsino mass $\mu$, and $\tan\beta$, assuming that all other supersymmetric particles and heavy Higgs bosons are decoupled.
As we will see, the ``blind spot''~\cite{1211.4873} of the Higgs-neutralino coupling plays an important role also in the Higgs- and $Z$-resonant neutralino DM scenario.
It is shown that there is still a large viable parameter space, and almost all the parameter space of the scenario will be covered complementarily by the future experiments.
Our results are summarized in Figs.~\ref{fig:main_tb2and3}--\ref{fig:main_tb3040and50}.

\section{Model}
\label{sec:model}
We assume that all SUSY particles but Bino and Higgsino multiplets are (much) heavier than 1 TeV, and the only scalar particle at the electroweak scale is the SM-like Higgs boson with the mass of 125 GeV.\footnote{
In the minimal SUSY SM, the Higgs mass is raised by the stop-loop~\cite{Higgs-stop} and the 125 GeV Higgs mass implies stop mass of ${\cal O}(1$--$10)~\TeV$ or heavier, depending on $\tan\beta$ and the $A$-term~\cite{Higgs-recent}.
For simplicity, we set the Higgs mass 125GeV and do not consider the heavier particles' effects.
} 
In this limit, the Wino component is decoupled, and the neutralino mass matrix becomes a $3\times 3$ matrix,
\begin{align}
M_\chi &=
\begin{pmatrix}
M_1 & -m_Z s_W \cos\beta & m_Z s_W \sin\beta \\
-m_Z s_W \cos\beta & 0 & -\mu \\
m_Z s_W \sin\beta &  -\mu & 0
\end{pmatrix}\,,
\label{eq:massmatrix}
\end{align}
in the basis of Bino and down-type and up-type Higgsinos, $(\widetilde{B}, \widetilde{H}^0_d, \widetilde{H}^0_u)$.
Here, $M_1$ and $\mu$ are the Bino and Higgsino masses, respectively, $\tan\beta=\vev{H_u}/\vev{H_d}$ is the ratio of the vacuum expectation values of the up- and down-type Higgs, and $s_W=\sin\theta_W$.
In this work, we assume that there is no CP-violation in the neutralino sector, and take $M_1>0$ and $\mu=$ real.
Then, the mass matrix is diagonalized by a real orthogonal matrix $O_\chi$ as
\begin{align}
O_\chi  M_\chi O_\chi^T
&=\text{diag}\left(\epsilon_1 m_{\chi_1}, \epsilon_2 m_{\chi_2}, \epsilon_3 m_{\chi_3}\right)\,,
\label{eq:massandmixing}
\end{align}
where $\epsilon_i=\pm 1$ and $0<m_{\chi_1}<m_{\chi_2}<m_{\chi_3}$.
Analytical and approximate formulae for the masses $m_{\chi_i}$ and the mixing matrix $O_\chi$ are given in Appendix~\ref{app:massandmixing}. 
The lightest neutralino $\chi_1$ is the DM candidate.
In the chargino sector, there is only one light chargino with mass $m_{\chi^\pm} = |\mu|$.
We assume $m_{\chi^\pm}>100~\GeV$ to avoid the LEP bound~\cite{Agashe:2014kda}.

In the following, we only consider a light Bino,
\begin{align}
M_1 < 80~\GeV\,.
\end{align}
The lightest neutralino then becomes Bino-like, and its coupling to the SM is given by the following Lagrangian,
\begin{align}
{\cal L}_{\chi_1\text{-SM}}
&=\frac{1}{2}\lambda^h h\;\overline{\psi_1}\psi_1
-\frac{1}{2}\lambda^Z Z_\mu \overline{\psi_1}\gamma^\mu\gamma^5 \psi_1\,,
\label{eq:DM-SM}
\end{align}
where $h$, $Z_\mu$ and $\psi_1$ denote the fields of the the SM-like Higgs boson, $Z$-boson, and the lightest neutralino, respectively.
The DM thermal relic abundance, its direct and indirect detections, and the Higgs and $Z$ invisible decay rates are all determined by the DM mass $m_{\chi_1}$ and the two couplings $\lambda^h$ and $\lambda^Z$.
The couplings are given by\footnote{
We have used the relation in the decoupling limit of the Higgs sector, $\alpha\simeq \beta-\pi/2$ with $\alpha$ being the mixing angle of the Higgs sector.
We also neglect radiative corrections by heavier particles.
}
\begin{align}
\lambda^h &= 
g'\epsilon_1 \OiB{1} \left( \OiHd{1}\cos\beta - \OiHu{1}\sin\beta \right)\,,
\label{eq:lambdaH} \\
\lambda^Z &= 
\frac{1}{2} g_Z \left( - [\OiHd{1}]^2 + [\OiHu{1}]^2 \right)\,,
\end{align}
where $g_Z=g/\cos\theta_W=g'/\sin\theta_W$, and $g$ and $g'$ are the SU(2)$_L$ and U(1)$_Y$ gauge couplings, respectively.

In terms of ${\cal O}(m_Z s_W / \mu)$ expansion, they are approximately given by (cf. Appendix~\ref{app:massandmixing})
\begin{align}
\lambda^h &\simeq
g' \epsilon_1 \left( \frac{\mu \sin 2\beta + M_1}{\mu^2-M_1^2} m_Z s_W + {\cal O} \lrfp{m_Z s_W}{\mu}{3} \right)\,,
\label{eq:lambdaHapprox}\\
\lambda^Z &\simeq 
\frac{1}{2}g_Z \left( \cos 2\beta \frac{m_Z^2 s_W^2}{\mu^2-M_1^2} + {\cal O} \lrfp{m_Z s_W}{\mu}{4} \right)\,.
\label{eq:lambdaZapprox}
\end{align}
From Eq.~\eqref{eq:lambdaHapprox}, the DM-$h$ coupling vanishes when
\begin{align}
M_1\simeq -\mu\sin 2\beta\,.
\label{eq:blindspot}
\end{align}
This leads to a ``blind spot"~\cite{1211.4873}, where the Higgs resonant annihilation, the spin-independent DM scattering for the direct detection, and the Higgs invisible decay are all suppressed.\footnote{
In Ref.~\cite{1211.4873}, all blind spots relevant to spin-independent and spin-dependent scattering are identified, and their phenomenological implications are investigated in the region of heavier neutralino DM, $m_{\chi_1}\gtrsim m_W$.
In the present scenario, only the one of Eq.~\eqref{eq:blindspot} is realized among those blind spots.
In this work, we do not consider the case of $\tan\beta \simeq 1$, which would lead to another blind spot for SD-scattering. The result for $\tan\beta \simeq 1$ will be similar to the case of $\tan\beta=2$ except for the SD scattering.
}
This suppression of $\lambda^h$ results in a parameter region that is not probed by the direct detection nor the Higgs invisible decay, which is important especially for the $Z$-resonant region $m_{\chi_1}\sim m_Z/2$.
On the other hand, the DM-$Z$ coupling is almost independent of $\tan\beta$ for $\tan\beta\gg 1$, and has only a mild dependence on $\tan\beta$ as far as $\tan\beta\gtrsim 2$. 

At the LHC, searches for charginos and neutralinos can probe this scenario, which will be discussed in Sec.~\ref{subsec:LHC}. 
The relevant interaction Lagrangian is given by
\begin{align}
{\cal L}&= W_\mu^- \sum_{i=1}^3 \overline{\psi_i}\gamma^\mu \left( \lambda_{LiC}^{W} P_L + \lambda_{RiC}^{W} P_R \right) \psi_C +\text{h.c.}
\nonumber\\
&+ Z_\mu \sum_{i<j} \overline{\psi_i}\gamma^\mu \left( \lambda_{Lij}^{Z} P_L + \lambda_{Rij}^{Z} P_R \right) \psi_j
\nonumber\\
&+ h \sum_{i<j} \overline{\psi_i} \left( \lambda_{Lij}^{h} P_L + \lambda_{Rij}^{h} P_R \right) \psi_j\,,
\label{eq:LforLHC}
\end{align}
where
\begin{align}
\lambda_{LiC}^{W}& = -\frac{1}{\sqrt{2}}g\eta_i \OiHu{i} \,,
\label{eq:LiCW} \\
\lambda_{RiC}^{W}&= \text{sign}(\mu) \frac{1}{\sqrt{2}}g\eta_i^* \OiHd{i} \,,
\label{eq:RiCW}\\
\lambda_{Lij}^Z = - (\lambda_{Rij}^Z)^* &= \frac{1}{2}g_Z \eta_i\eta_j^* \left( - \OiHd{i}\OiHd{j} + \OiHu{i}\OiHu{j} \right) \,,
\label{eq:LijZ}\\
\lambda_{Lij}^{h} = (\lambda_{Rij}^{h})^* &= \frac{1}{2}g'\eta_i^*\eta_j^* \left[ \OiB{i} \left( \OiHd{j}\cos\beta - \OiHu{j}\sin\beta \right) + (i\leftrightarrow j) \right]\,.
\label{eq:Rijh}
\end{align}
Here, $\eta_i^2=\epsilon_i$, and $\psi_C$ denotes the chargino field which is defined to have a mass term $-{\cal L}=\text{sign}(\mu)\mu \overline{\psi_C}\psi_C$ and to have a positive charge.

\section{Constraints and Prospects}
Our main results are shown in Figs.~\ref{fig:main_tb2and3}--\ref{fig:main_tb3040and50}, where the constraints and prospects listed in Sec.~\ref{sec:intro} are presented in the $(m_{\chi_1}, m_{\chi^\pm})$-planes for $2 \leq \tan\beta \leq 50$.
In the figures, we show only the region with $30~\GeV \le m_{\chi_1}\le 70~\GeV$, because the relic density is always too large outside this region for $m_{\chi^\pm}>100~\GeV$ and $M_1<80~\GeV$.
In the following subsections, we explain each of the constraints and prospects in turn.
We also briefly mention other possible constraints in Sec.~\ref{subsec:others}.

\subsection{Thermal relic abundance}
\label{subsec:relic}
We assume that the present energy density of DM is dominantly given by that of the thermal relic of the lightest neutralino. 
In the present scenario, the lightest neutralino can only annihilate into a pair of SM fermions, and the annihilation cross section is given by
\begin{align}
\sigma(\chi_1\chi_1\to f\bar{f}) = \sigma(\chi_1\chi_1\to h^*\to f\bar{f}) + \sigma(\chi_1\chi_1\to Z^*\to f\bar{f})\,,
\end{align}
where
\begin{align}
\sigma(\chi_1\chi_1\to h^*\to f\bar{f})
&\simeq
\frac{1}{2} (\lambda^h)^2 \sqrt{1-\frac{4m_{\chi_1}^2}{s}} \frac{1}{(s-m_h^2)^2+(m_h\Gamma_h)^2} \frac{s}{m_h} \Gamma(h\to f\bar{f})\,,
\label{eq:annihilationH} \\
\sigma(\chi_1\chi_1\to Z^*\to f\bar{f})
&\simeq
(\lambda^Z)^2 \sqrt{1-\frac{4m_{\chi_1}^2}{s}} \frac{1}{(s-m_Z^2)^2+(m_Z\Gamma_Z)^2} \frac{s}{m_Z} \Gamma(Z \to f\bar{f})\,,
\label{eq:annihilationZ}
\end{align}
and there is no interference term.
Here, we have neglected the terms proportional to $(m_f/m_{\chi_1})^2\ll 1$, where $m_f$ is the mass of final state fermion.
The DM abundance is calculated by solving the Boltzmann equation
\begin{align}
\frac{dY_{\chi_1}}{dt} = -n_S\vev{\sigma v_{\text{rel}}}(Y_{\chi_1}^2-Y_{\chi_1,\text{eq}}^2)\,,
\end{align}
where the thermal average of the annihilation cross section times the relative velocity (with Maxwell-Boltzmann equilibrium distribution) is given by~\cite{Gondolo:1990dk},
\begin{align}
\vev{\sigma v_{\text{rel}}}(T)
&= \frac{ \int d^3p_1 d^3p_2\; e^{-E_1/T}e^{-E_2/T} \sigma  v_{\text{rel}} }{ \int d^3p_1 d^3p_2\; e^{-E_1/T}e^{-E_2/T} }
\nonumber\\
&= \frac{1}{8m_{\chi_1}^4T [K_2(m_{\chi_1}/T)]^2} \int^\infty_{4m_{\chi_1}^2}\sigma (s)\sqrt{s}(s-4m_{\chi_1}^2)K_1(\sqrt{s}/T) ds\,.
\end{align}
Here, $Y_{\chi_1\,(\text{eq})}=n_{\chi_1\,(\text{eq})}/n_S$, $n_S=(2\pi^2/45)g_{*S}T^3$ is the entropy density, $n_{\chi_1}$ is the DM number density, $n_{\chi_1,\text{eq}}=2(m_{\chi_1}^2 T/2\pi^2)K_2(m_{\chi_1}/T) \simeq 2(m_{\chi_1}T/2\pi)^{3/2}\exp(-m_{\chi_1}/T)$ is its equilibrium value, and $t$ and $T$ are the cosmic time and the temperature, respecitvely.\footnote{
$T$ and $t$ are related by $dt/dT=(HT)^{-1}[1+(1/3)d(\ln g_{*S})/d(\ln T)]$, where $H=(\pi^2 g_*/90)^{1/2}T^2/M_P$ is the Hubble parameter with the reduced Planck scale $M_P\simeq 2.44\times 10^{18}~\GeV$.
For the effective degrees of freedom $g_*(T)$ and $g_{*S}(T)$, we have used the fitting formula in~\cite{0910.1066}.
}
$K_{1,2}$ are the modified Bessel functions of the first and second kind. 
The final relic abundance is given by $\Omega_{\chi_1} = m_{\chi_1}Y_{\chi_1}/(\rho_c/s)_0 $, where $(\rho/s)_0 \simeq 3.65 h^2\times 10^{-9}~\GeV$ is the critical density divided by the entropy density at present, with $h\simeq 0.67$ being the scale factor for Hubble constant~\cite{Agashe:2014kda}.\footnote{
We have also calculated the DM abundance with micrOMEGAs~\cite{micrOMEGAs}, and checked that the results agree within a few \%.
}

In Figs.~\ref{fig:main_tb2and3}--\ref{fig:main_tb3040and50}, the contours of relic DM density $\Omega_{\chi_1}h^2=0.120$ is shown in black lines.
The contours have clear peaks at the $Z$- and Higgs-resonances, $m_{\chi_1}\sim m_Z/2\simeq 45~\GeV$ and $m_{\chi_1}\sim m_h/2\simeq 62~\GeV$.
In these regions, the chargino mass $m_{\chi^\pm}$ can take a large value, corresponding to small DM-$Z$ and DM-Higgs couplings, $\lambda^Z$ and $\lambda^h$.

In the $Z$-resonant region, $m_{\chi_1}\sim m_Z/2$, the relic abundance shows a universal behavior for all $\tan\beta\gg 1$.
This is because the DM-$Z$ coupling $\lambda^Z$ is almost independent of $\tan\beta$ for $\tan\beta\gg 1$, as shown in Eq.~\eqref{eq:lambdaZapprox}.
In this region, the chargino mass $m_{\chi^\pm}$ is always bounded from above as $m_{\chi^\pm}\lesssim 450~\GeV$~\cite{1410.5730}, which corresponds to $|\lambda^Z|\gtrsim 0.0034$ [cf. Eq.~\eqref{eq:lambdaZapprox}].\footnote{
This is consistent with the analysis of generic $Z$-portal DM in Ref.~\cite{1411.2985}.
} 
This upper bound on the chargino mass is crucial for the LHC search discussed in Sec.~\ref{subsec:LHC}.

In the Higgs resonant region, $m_{\chi_1}\sim m_h/2$, the behavior of the relic abundance in the $(m_{\chi_1},m_{\chi^\pm})$-planes strongly depends on $\tan\beta$ as well as the sign$(\mu)$.
This  is also understood in terms of the DM-Higgs coupling, $\lambda^h$, in Eq.~\eqref{eq:lambdaHapprox}.
As can be seen in Figs.~\ref{fig:main_tb45and6}, for $\mu<0$ and $4\lesssim \tan\beta \lesssim 6$, there are two regions corresponding to $\Omega_{\chi_1}h^2\le 0.12$.
This is because of the blind-spot behavior discussed in Sec.~\ref{sec:model}.
The coupling $\lambda^h$ has  opposite signs in the two separate regions, and it becomes zero in between.
For $\mu<0$ and $\tan\beta\gtrsim 7$, the region of large $m_{\chi^\pm}$ disappears because a sufficiently large $|\lambda^h|$ can no longer be obtained there.
For both $\mu<0$ and $\mu>0$ and for all $\tan\beta$, the maximal chargino mass corresponds to $|\lambda^h| \simeq 0.0052$.
For $\tan\beta\gtrsim 10$, the upper bound on the chargino mass is as small as $m_{\chi^\pm}\lesssim 400~\GeV$ for $\mu<0$ and $m_{\chi^\pm}\lesssim (500$--$800)~\GeV$ for $\mu>0$.
As we shall see in Sec.~\ref{subsec:LHC}, these regions can be probed by the 14 TeV LHC.
For small $\tan\beta$, however, a much larger chargino mass is allowed, e.g., $m_{\chi^\pm}\lesssim 2.5~\TeV$ for $\tan\beta=2$ and $\mu>0$.
Although such a heavy chargino is out of the 14 TeV LHC reach, the direct detection experiments can cover most of the region, as we will see in the next subseciton.

\subsection{Direct detection}
\label{subsec:direct}
In the present scenario, the spin-independent (SI) and spin-dependent (SD) scatterings between the DM and nuclei are induced by Higgs-exchange and $Z$-exchange, respectively.
As we shall see, the former gives a strong bound and high future sensitivity, while the latter plays a complementary role in the blind spot regions for $\mu<0$.

\subsubsection{spin-independent scattering}
The SI scattering cross section of DM per nucleon is given by
\begin{align}
\sigma^{\text{SI}}_N &=
\frac{4}{\pi}\lambda_N^2 m_N^2 \left(1+\frac{m_N}{m_{\chi_1}}\right)^{-2}\,,
\label{eq:SIXS}
\end{align}
where $m_N$ is the nucleon mass and $\lambda_N$ is the effective coupling between the DM and nucleons, ${\cal L}=\sum_{N=p,n} \lambda_N\overline{\psi_1}\psi_1 \overline{N} N$.
In our scenario, the coupling is induced by the Higgs exchange, and given by~(cf.~\cite{Shifman:1978zn,micrOMEGAs})
\begin{align}
\lambda_N &= \frac{\lambda^h}{2m_h^2} \cdot \frac{m_N f_N}{2m_W/g}\,,
\quad
f_N= \sum_{q=u,d,s} f_{T_q}^{(N)} + \frac{2}{9} \left(1-\sum_{q=u,d,s} f_{T_q}^{(N)}\right)\,,
\end{align}
where $f_{T_q}^{(N)}=\vev{N|m_q \bar{q}q|N}/m_N$.
In our analysis, we use the default values of $f_{T_q}^{(p)}$ adopted in micrOMEGAs 4.1~\cite{micrOMEGAs}, $f_{T_u}^{(p)}=0.0153$, $f_{T_d}^{(p)}=0.0191$, and $f_{T_s}^{(p)}=0.0447$, which leads to $f_N = 0.284$.\footnote{
Note that the strange quark mass fraction $f_{T_s}^{(p)}$ is already small.
If we use the result of the lattice calculation in Ref.~\cite{1208.4185}, which is further smaller as $f_{T_s}^{(p)}\simeq 0.009$, it leads to $f_N\simeq 0.256$, resulting in 20\% smaller SI scattering cross sections.
We have also calculated the SI cross section with micrOMEGAs 4.1~\cite{micrOMEGAs}, and checked that the results agree within a few \%.
}
Therefore, the SI scattering cross section is given by
\begin{align}
\sigma^{\text{SI}}_N \simeq
5.2\times 10^{-43}\cdot (\lambda^h)^2 \left(1+\frac{m_N}{m_{\chi_1}}\right)^{-2}~\text{cm}^2\,.
\end{align}

In Fig.~\ref{fig:main_tb2and3}--\ref{fig:main_tb3040and50}, we show the constraint obtained by the LUX~\cite{1310.8214} (90\% CL limit) and the future prospects for the XENON 1T~\cite{1206.6288} with blue dashed and solid lines, respectively.
The results are understood in terms of the coupling $\lambda^h$ in Eqs.~\eqref{eq:lambdaH} and \eqref{eq:lambdaHapprox}.
As can be seen in the figures, for $\mu>0$, the region with $m_{\chi^\pm}\lesssim 120$--400~GeV (150--400~GeV) are excluded by the LUX, for $m_{\chi_1}\sim m_Z$ ($m_{\chi_1}\sim m_h$).
The XENON 1T can cover most of the viable parameter space for $\mu>0$, except for the peak of the Higgs-resonance, where $\lambda^h\simeq 0.0052$ (cf. Sec.~\ref{subsec:relic}) and $\sigma^{\text{SI}}_N\simeq 1.4\times 10^{-47}\text{cm}^2$, and the $Z$ peaks for $\tan\beta\gtrsim 30$.
Note that the SI cross sections in these peak regions are just below the sensitivity shown in Ref.~\cite{1206.6288}.
Therefore, it is expected that future experiments with higher sensitivity~\cite{1310.8327} can cover the whole parameter region for $\mu>0$.

For $\mu<0$, because of the cancellation in Eq.~\eqref{eq:lambdaHapprox}, the constraint and  the sensitivity are significantly reduced in terms of the chargino mass $m_{\chi^\pm}$.
In the Higgs-resonant region, the parameter regions with the correct thermal relic abundance, $\Omega_{\chi_1}h^2\simeq 0.12$, will still be mostly covered by the XENON 1T.
This is because both of $\Omega_{\chi_1}h^2$ and $\sigma^{\text{SI}}_N$ are determined by the same coupling, $\lambda^h$.
The correlation is clearly seen, e.g., in Fig.~\ref{fig:main_tb45and6}, for $5\lesssim \tan\beta\lesssim 6$.
In the $Z$-resonant region, however, $\Omega_{\chi_1}h^2$ and $\sigma^{\text{SI}}_N$ are determined by different couplings, $\lambda^Z$ and $\lambda^h$, respectively.
This results in a large parameter region which gives correct $\Omega_{\chi_1}h^2$ but very small $\sigma^{\text{SI}}_N$, as can be seen in Figs.~\ref{fig:main_tb78and9}--\ref{fig:main_tb3040and50}.
Since $\lambda^h$ can be zero at a certain value of $\mu$, there always remains a region which cannot be probed by the SI scattering.
Some of these regions are probed by SD scattering discussed in the next subsection, and the search for the chargino and the neutralino at the LHC, discussed in Sec.~\ref{subsec:LHC}, will be a very sensitive probe in these regions.

\subsubsection{spin-dependent scattering}
Now let us discuss the SD scattering.
The SD nucleon-DM scattering cross section is given by~\cite{hep-ph/9506380,micrOMEGAs}
\begin{align}
\sigma^{\text{SD}}_N & = \frac{12}{\pi}\xi_N^2m_N^2 \left(1+\frac{m_N}{m_{\chi_1}}\right)^{-2}\,,
\end{align}
where $\xi_N$ is the axial-vector effective coupling between the DM and the nucleons, ${\cal L}=\sum_{N=p,n} \xi_N\overline{\psi_1}\gamma^5\gamma_\mu\psi_1 \overline{N}\gamma^5\gamma^\mu N$.
In our scenario, the coupling is induced by the $Z$ exchange, and given by~(cf.~\cite{hep-ph/9506380,micrOMEGAs})
\begin{align}
\xi_N = \frac{\lambda^Z g_Z}{4m_Z^2}\sum_{q=u,d,s} T^3_q \Delta^{(N)}_q\,,
\end{align}
where $\Delta_q^{(N)}$ parametrize the quark spin content of the nucleon.
In our analysis, we use the default values adopted in micrOMEGAs~\cite{micrOMEGAs}, $\Delta_u^{(p)}=\Delta_d^{(n)}=0.842$, $\Delta_d^{(p)}=\Delta_u^{(n)}=-0.427$, and $\Delta_s^{(p)}=\Delta_s^{(n)}=-0.085$.
Therefore, the SD nucleon-DM scattering cross section is given by\footnote{
We have again checked the SD cross section with micrOMEGAs  and the results agree within 1 \%.
}
\begin{align}
\sigma^{\text{SD}}_{n (p)}
&\simeq 2.3 (3.0)\times 10^{-37}\cdot (\lambda^Z)^2 \left(1+\frac{m_{n (p)}}{m_{\chi_1}}\right)^{-2} \;\text{cm}^2\,.
\label{eq:SDXS}
\end{align}

In Fig.~\ref{fig:main_tb2and3}--\ref{fig:main_tb3040and50}, we show the constraint on the SD neutron-DM scattering cross section $\sigma^{\text{SD}}_n$ from the XENON100~\cite{1301.6620} with green dashed lines.\footnote{
See also Ref.~\cite{1211.4573}.
The constraint on proton-DM scattering, $\sigma^{\text{SD}}_p$, is much weaker.
The LUX has not published constraints on SD scattering cross section.
If we adopt the constraint on SD scattering cross section in Ref.~\cite{1407.8257}, which is based on the LUX data, the bound on the chargino mass becomes about 20--30\% stronger.
}
As can be seen from the figures, this bound gives the strongest current constraint in some of the small $m_{\chi^\pm}$ regions for $\mu<0$.
The bound on the chargino is about $m_{\chi^\pm}\lesssim 100$--140~GeV, and it has only very mild dependences on  $\tan\beta$, $m_{\chi_1}$, and sign$(\mu)$.
This can be understood from Eqs.~\eqref{eq:lambdaZapprox} and \eqref{eq:SDXS}, which lead to the following approximate formula,
\begin{align}
\sigma^{\text{SD}}_{n (p)}
&\simeq 1.4 (1.8)\times 10^{-41} \lrfp{300~\GeV}{|\mu|}{4} \left(1-\frac{m_{\chi_1}^2}{\mu^2}\right)^{-2} (\cos 2\beta)^2\;\text{cm}^2\,.
\end{align}
In the figures, we also show the prospect of the XENON 1T for the $\sigma^{\text{SD}}_n$, studied in Ref.~\cite{1211.4573}, with green solid lines.
It reaches the chargino mass of about 280--350~GeV, which will cover a large part of the blind spot for $\mu<0$.

\subsection{Higgs and $Z$ invisible decays}
For $m_{\chi_1}<m_h/2$ the Higgs boson can decay into a pair of DMs, which lead to a invisible decay.
The branching ratio is given by
\begin{align}
\text{Br}(h\to \chi_1\chi_1) 
&= \frac{\Gamma(h\to \chi_1\chi_1)} {\Gamma(h\to\text{SM})+\Gamma(h\to \chi_1\chi_1)}\,,
\end{align}
where
\begin{align}
\Gamma(h\to \chi_1\chi_1)=\frac{(\lambda^h)^2}{16\pi} m_h \left(1-\frac{4m_{\chi_1}^2}{m_h^2} \right)^{3/2}\,.
\end{align}
We have used $\Gamma(h\to\text{SM})=4.07~\MeV$~\cite{Agashe:2014kda} in our calculation. 
The constraint on the Higgs invisible decay has been obtained by global fits to Higgs data~\cite{1303.3570,1306.2941}.
In our numerical calculation, we adopt the one in Ref.~\cite{1306.2941}, $\text{Br}(h\to \text{invisible})<0.19$ (95\% CL).

As for the future prospects, we consider the high-luminosity (HL) LHC and the ILC.
The sensitivity of the HL-LHC depends on the systematic uncertainties.
Here, we use the estimated future sensitivity of searches for Higgs decaying invisibly using $ZH$ channel in Ref.~\cite{1309.7925}, $\text{Br}(h\to \text{invisible})<0.062$ (95\% CL) for 3000~fb$^{-1}$, adopting the value of the ``realistic scenario'' for the size of systematics.
In the ``conservative scenario'' the estimated sensitivity becomes $\text{Br}(h\to \text{invisible})<0.14$.
In the Higgs working group report of the 2013 Snowmass~\cite{1310.8361}, the 95\% CL limit (for 3000~fb$^{-1}$) is estimated as $\text{Br}(h\to \text{invisible})<0.08$--0.16 for ATLAS and 0.06--0.17 for CMS.
For the ILC, we use the value in~\cite{1310.0763}, $\text{Br}(h\to \text{invisible})<0.004$ (1150 fb$^{-1}$ at $\sqrt{s}=250~\GeV$).

The constraint and prospects for the Higgs invisible decay are shown in Fig.~\ref{fig:main_tb2and3}--\ref{fig:main_tb3040and50} with magenta lines.
As can be seen in the figures, a large parameter space is covered by the Higgs invisible decay search.
For $\mu>0$, the whole $Z$-resonant region will be covered by the ILC.
The blind spots are again clearly seen for $\mu<0$, where Higgs invisible decay becomes very small due to the suppression of $\lambda^h$ in Eq.~\eqref{eq:lambdaHapprox}.

Before closing this subsection, let us comment on the $Z$ invisible decay.
For $m_{\chi_1}<m_Z/2$, the $Z$ can decay into a pair of DMs, with a partial decay rate
\begin{align}
\Gamma(Z\to \chi_1\chi_1)=\frac{(\lambda^Z)^2}{24\pi} m_Z \left(1-\frac{4m_{\chi_1}^2}{m_Z^2} \right)^{3/2}\,.
\end{align}
The bound from the LEP, $\Gamma(Z\to \chi_1\chi_1)<2.0~\MeV$ (95\% CL)~\cite{hep-ex/0509008}, corresponds to $\lambda^Z\lesssim 0.041\times (1-4m_{\chi_1}^2/m_Z^2)^{-3/2}$.
In our setup, this is always weaker than the bound on $\sigma^{\text{SD}}$ from the XENON100~\cite{1301.6620} for $m_{\chi_1}\gtrsim 30~\GeV$~(see Sec.~\ref{subsec:direct}), and hence we do not show the bound in the figures.

\subsection{Search for the chargino and neutralinos at the LHC}
\label{subsec:LHC}
As we have seen in Sec.~\ref{subsec:relic}, the requirement that the thermal relic abundance of $\chi_1$ explains the observed DM density, $\Omega_{\chi_1}h^2\simeq 0.12$, gives upper bounds on the chargino mass $m_{\chi^\pm}$ in the present scenario.
The chargino mass is ${\cal O}(100~\GeV)$ except for the Higgs resonance peak for small $\tan\beta$.
In particular, in the $Z$-resonant region, the chargino mass $m_{\chi^\pm}$ is always bounded from above as $m_{\chi^\pm}\lesssim 450~\GeV$~\cite{1410.5730}.
The heavier neutralinos, $\chi_2$ and $\chi_3$, also have masses $m_{\chi_{2,3}}\simeq m_{\chi^\pm}$.
These ${\cal O}(100~\GeV)$ chargino and neutralinos are within the reach of the LHC experiments.

In this work, we consider the following production and decay channels at the LHC,
\begin{eqnarray}
pp\to \chi_{2,3}\chi^\pm \to Z\chi_{1}\;W^\pm\chi_{1} \to \ell \ell \chi_{1}\; \ell \nu\chi_{1}\,,
\label{eq:WZatLHC}
\end{eqnarray}
which leads to a signal with three leptons and missing energy, and gives a high sensitivity in the present scenario~\cite{1409.6322}.
The sensitivities of the SUSY searches at 14 TeV, including the high luminosity run of $3000~\text{fb}^{-1}$, is studied by ATLAS~\cite{ATL-PHYS-PUB-2014-010} and CMS~\cite{CMS:2013ega}.
In this work, we reinterpret the ATLAS study~\cite{ATL-PHYS-PUB-2014-010} of the search for the channel in Eq.~\eqref{eq:WZatLHC}, for $300~\text{fb}^{-1}$ and $3000~\text{fb}^{-1}$.
The constraints from the LHC run I~\cite{1402.7029} are discussed in Appendix~\ref{app:8TeV}.

In the ATLAS analysis~\cite{ATL-PHYS-PUB-2014-010}, the results for the three-lepton process~\eqref{eq:WZatLHC} is presented assuming a simplified ``pure Wino" model with three states; two neutralinos $\chi_1$, $\chi_2$ and one chargino $\chi^\pm$, with degenerate masses of the chargino and the heavier neutralino, $m_{\chi^\pm}=m_{\chi_2}$ and 100\% branching ratios of $\text{Br}(\chi_2\to \chi_1 Z)=1$ and $\text{Br}(\chi^\pm \to \chi_1 W)=1$.
There are several differences in the present setup.
(i) There are two heavier neutralinos $\chi_2$ and $\chi_3$.
(ii) The sum of the production cross sections of chargino-neturalino pair is a factor $\simeq 1/2$ smaller, since the ATLAS analysis assumes Wino-like chargino-neutralino pair production.
(iii) The neutralinos $\chi_2$ and $\chi_3$ have generically sizable branching fractions to the Higgs as well, $\text{Br}(\chi_{2,3}\to \chi_1 h)={\cal O}(1)$ and $\text{Br}(\chi_{2,3}\to \chi_1 Z)={\cal O}(1)$.
(iv) Although their masses $m_{\chi_{2,3}}$ are close to the chargino mass $m_{\chi^\pm}$, the difference can have a non-negligible effect on the production cross section, especially for small $|\mu|$.
Therefore, the results in Ref.~\cite{ATL-PHYS-PUB-2014-010} cannot be directly applied to the present model and reinterpretations are necessary.

In the ATLAS analysis, several signal regions (SRs) are defined by various kinematical cuts.
We calculate the expected number of events in each signal region (SR) X as 
\begin{align}
N_{\text{SR-X}}
= \sum_{j=2,3} & \sum_{\chi^\pm} \sigma^{\text{NLO}}(pp\to \chi^\pm\chi_j) \cdot \text{Br}(\chi^\pm\to\chi_{1}W^{(*)}\to\chi_1\ell\nu) \cdot \text{Br}(\chi_j\to\chi_{1}Z^{(*)}\to\chi_1\ell\ell)
\nonumber\\
&\times A_{\text{SR-X}} \cdot \int {\cal L}dt\,,
\label{eq:NSRX}
\end{align}
where $\int {\cal L}dt$ denote the integrated luminosity, and $A_{\text{SR-X}}$ is defined by
\begin{align}
A_{\text{SR-X}}
&= \frac{ \text{\# of events which pass the cuts of SR-X} }{ \text{\# of generated events in } pp\to \chi^\pm \chi_i\to W^{(*)} \chi_1 Z^{(*)}  \chi_1 \to \ell\nu \chi_1 \ell\ell \chi_1 }\,,
\label{eq:acc2}
\end{align}
where $\ell$ denotes $e$, $\mu$ and $\tau$.
For simplicity, we discard the hadronic decays of $W$ and $Z$.
The branching fractions are given by (when kinematically allowed)
\begin{align}
\text{Br}(\chi^\pm\to\chi_{1}W\to\ell\nu) 
&= \text{Br}(\chi^\pm\to\chi_{1}W) \cdot \text{Br}(W\to\ell\nu)\,, \\
\text{Br}(\chi_j\to\chi_{1}Z\to\ell\ell)
&= \text{Br}(\chi_j\to\chi_{1}Z) \cdot \text{Br}(Z\to\ell\ell)\,,
\end{align}
where the chargino branching fraction is $\text{Br}(\chi^\pm\to\chi_{1}W)=1$, while the neutralino branching is given by
\begin{align}
\text{Br}(\chi_{j}\to Z\chi_1)
&= \frac{\Gamma_{j}(\chi_{j}\to Z\chi)}{\Gamma_{j}(\chi_{j}\to Z\chi)+\Gamma_{j}(\chi_{j}\to h\chi)}\,, 
\label{eq:BranchingRatio}\\
\Gamma(\chi_{j}\to Z\chi)
&= \frac{1}{16\pi}m_{\chi_j}|\lambda_{L1j}^Z|^2 \left( 1+6\epsilon_1\epsilon_{j} r^1_{j}+(r^1_{j})^2-2(r^z_{j})^2+\frac{(1-(r^1_{j})^2)^2}{(r^z_{j})^2} \right)
\nonumber\\
&\times \left(1-(r^1_{j}-r^z_{j})^2\right)^{1/2}\left(1-(r^1_{j}+r^z_{j})^2\right)^{1/2}\,, \\
\Gamma(\chi_{j}\to h\chi)
&= \frac{1}{16\pi}m_{\chi_j}|\lambda_{L1j}^h|^2 \left( 1+2\epsilon_1 \epsilon_{j} r^1_{j} +(r^1_j)^2-(r^h_{j})^2 \right)
\nonumber\\
&\times  \left(1-(r^1_{j}-r^h_{j})^2\right)^{1/2}\left(1-(r^1_{j}+r^h_{j})^2\right)^{1/2}\,,
\end{align}
where $r^1_{j}=m_{\chi_1}/m_{\chi_j}, r^z_{j}=m_Z/m_{\chi_j}, r^h_{j}=m_h/m_{\chi_j}$, and the couplings are given by Eqs.~\eqref{eq:LijZ} and \eqref{eq:Rijh}.

In the numerical calculations, we generate the events with MadGraph5\_aMC@NLO 2.2.3~\cite{1405.0301} in combination with PYTHIA 6.4~\cite{hep-ph/0603175}.
We generate the events at LO and rescale the acceptance with the NLO cross section calculated by Prospino 2.1~\cite{prospino} with CTEQ6L1~\cite{hep-ph/0201195} parton distribution functions (PDFs).
Delphes 3~\cite{1307.6346} is used with the ATLAS parameters card\footnote{
$b$-tagging efficiencies and lepton isolation criteria are arranged as used in the ATLAS analysis~\cite{ATL-PHYS-PUB-2014-010}.
}
given in the MadGraph5\_aMC@NLO package for the fast detector simulation.

There are three (four) SRs considered to probe the signal of Eq.~\eqref{eq:WZatLHC} for 300 (3000) fb$^{-1}$, denoted as SRA--SRC (SRA--SRD).
In our analysis, after electrons, muons, and jets are selected following the ATLAS analysis~\cite{ATL-PHYS-PUB-2014-010},
the following cuts are applied.
\begin{itemize}
\item There should be exactly three leptons in each event, and at least one SFOS lepton pair is required to have invariant mass $|m_{\text{SFOS}}-M_Z|<10~\GeV$.
\item Events with $b$-tagged jets are discarded.
\item The three lepton $p_T$ should be larger than 50 GeV.
\item Then, the events are divided into SRs depending on the missing transverse energy $E_T^{\text{miss}}$ and the transverse mass $m_T$, where $m_T$ is calculated with the missing transverse energy and the lepton which does not form the SFOS lepton pair whose mass is closest to the $Z$-boson mass. 
\end{itemize}

As a validation of our analysis, we have calculated the expected numbers of events in each SRs for the ``pure Wino" model points with $(m_{\chi_2}, m_{\chi_1})=$ (400,0), (600,0), (800,0), (1000,0)~GeV, which are studied in the ATLAS analysis~\cite{ATL-PHYS-PUB-2014-010}. 
They are in good agreement with the ATLAS analysis.

In the ATLAS analysis~\cite{ATL-PHYS-PUB-2014-010}, the expected 95\% exclusion limit is shown in the $(m_{\chi_2},m_{\chi_1})$-plane by combining disjoint versions of SRs.
We have analyzed the same parameter space as a validation.
In our analysis, the expected exclusion line is obtained as follows.
 (i) For each SR, the expected upper limit on the number of beyond-the-SM events is calculated from the number of the background events given in~\cite{ATL-PHYS-PUB-2014-010}, using $Z_N$=1.64~\cite{Linnemann:2003vw} for 95\% CL exclusion.
(ii) At each model point, the expected number of signal events in each SR is calculated.
(iii) The model point is excluded if and only if it is excluded in at least one of the SRs. 
The obtained expected exclusion line, $m_{\chi^\pm} \lesssim$ 800 (1100) GeV for $m_{\chi_1} \lesssim 100~\GeV$ at 300 (3000) fb$^{-1}$, agrees with the ATLAS result within an error of $\delta m_{\chi^\pm}\simeq 20~\GeV$.

Next, we apply the same analysis in the present scenario, i.e., the Higgs- and $Z$-resonant neutralino DM.
The cross sections and acceptances are calculated as follows.
For $|\mu|=100, 110, \cdots 1000~\GeV$, the cross sections are calculated at $(\tan\beta,\text{sign}(\mu),M_1)=(2,+,80~\GeV)$ and $(50,+,30~\GeV)$.\footnote{These points are chosen since they give the smallest, medium and largest value of $m_{\chi_j}-m_{\chi^\pm}$.}
Then, the cross sections normalized by the coupling, $\sigma^{\text{NLO}}/(|\OiHu{j}|^2+|\OiHd{j}|^2)$, are interpolated.
The acceptances are calculated by varying $m_{\chi^\pm}$, $m_{\chi_j}-m_{\chi^\pm}$ and $m_{\chi_1}$ by $(\Delta m_{\chi^\pm}, \Delta (m_{\chi_j}-m_{\chi^\pm}), \Delta m_{\chi_1})=(20,10,5)~\GeV$ for 100$~\GeV<|\mu|<300~\GeV$ and $(\Delta m_{\chi^\pm}, \Delta (m_{\chi_j}-m_{\chi^\pm}), \Delta m_{\chi_1})=(20,10,10)~\GeV$ for  300$~\GeV<|\mu|<1000~\GeV$, while the couplings are fixed as the ones of $\tan\beta=5, M_1=50~\GeV$, $\mu=200~\GeV$, for simplicity.\footnote{
We have checked that the acceptance does not depend much on these parameters.
}
Here, we do not consider the region of $m_{\chi_j}-m_{\chi_1} < m_Z$, for simplicity.

Results are shown in Fig.~\ref{fig:main_tb2and3}--\ref{fig:main_tb3040and50} with red lines.
The expected exclusion region at 300~fb$^{-1}$ is shown in light orange region with red dotted lines.
One can see that the $Z$-peaks in the whole parameter space, including the blind spot, will be probed at 300~fb$^{-1}$.
For $\tan\beta\geq 30$, the Higgs peaks can also be covered.
The small $m_{\chi^\pm}$ region is not covered because of the small mass differences between $\chi_{2,3}, \chi^\pm$ and $\chi_1$.

At 3000 fb$^{-1}$, much larger parameter space will be probed, up to $m_{\chi^\pm}\sim 800~\GeV$.
The Higgs-resonant regions are covered for $\tan\beta\gtrsim 15$ ($\tan\beta\geq$ 6) for $\mu > 0$ ($\mu < 0$).
Though the small $m_{\chi^\pm}$ region can not be covered even at 3000 fb$^{-1}$, combination with other experiments such as the direct searches can probe almost all the parameter region of the present scenario.

As can be seen in the figures, the expected reach for the chargino mass, $m_{\chi^\pm}\sim 800~\GeV$, is almost independent of $\tan\beta$ and $m_{\chi_1}$. 
This can be understood as follows.
In the large $m_{\chi^\pm}$ region, the cross section is mainly determined by $|\mu|$ because the masses are almost degenerate as $m_{\chi_2}\simeq m_{\chi_3}\simeq m_{\chi^\pm}=|\mu|$, and the coupling with $W$-boson is universal for $|\mu|\gg m_Z$ [cf. Eqs.~\eqref{eq:LiCW},~\eqref{eq:RiCW},~\eqref{eq:OuHdj},~\eqref{eq:OuHuj}].
In addition, because of the large mass hierarchy $m_{\chi_{2,3}}\simeq m_{\chi^\pm}\gg m_{\chi_1}, m_Z, m_W$, the acceptance is determined almost only by $|\mu|$.
Thus, from Eq.~\eqref{eq:NSRX}, $N_{\text{SR-X}}$ becomes
\begin{eqnarray}
N_{\text{SR-X}}
\simeq &
\sum_{\chi^\pm} \sigma^{\text{NLO}}(pp\to \chi^\pm\chi_2) \cdot \text{Br}(\chi^\pm\to\chi_{1}W\to\chi_1\ell\nu)\cdot A_{\text{SR-X}} \cdot \int {\cal L}dt\nonumber
\\
&\times \sum_{j=2,3} \text{Br}(\chi_j\to\chi_{1}Z\to\chi_1\ell\ell)\,,
\end{eqnarray}
The first line of this equation is determined almost only by $|\mu|$.
The second line can be expanded in terms of ${\cal O}(m_Z s_W / \mu)$ as
\begin{eqnarray}
\text{Br}(\chi_2\to Z\chi_1)&=&\frac{1}{2}(1+\sin{2\beta})+\frac{M_1}{\mu}(1-\sin^2{2\beta})+{\cal O}\left(\frac{m_Z s_W}{\mu}\right)^2\,,\\
\text{Br}(\chi_3\to Z\chi_1)&=&\frac{1}{2}(1-\sin{2\beta})-\frac{M_1}{\mu}(1-\sin^2{2\beta})+{\cal O}\left(\frac{m_Z s_W}{\mu}\right)^2\,.
\end{eqnarray}
From this expression, $\text{Br}(\chi_2\to Z\chi_1)+\text{Br}(\chi_3\to Z\chi_1)\simeq 1$ for $|\mu|\gg m_Z$, and it is almost 
independent of $\tan\beta$ and $m_{\chi_1}$.

\subsection{Other constraints}
\label{subsec:others}
Let us briefly comment on other possible constraints on the present scenario.
\begin{itemize}
\item
indirect search.
\\
The DM annihilation in the present Universe can lead to cosmic rays such as photons, positrons, and anti-protons.
In the present model, however, the annihilation cross section in the present Universe is suppressed by the velocity, $\vev{\sigma v_{\text{rel}}}_0 \sim v^2$, as shown in Eqs.~\eqref{eq:annihilationH} and \eqref{eq:annihilationZ}.
In the limit of $v\to 0$, the leading term in the amplitude comes from the $Z$-exchange diagram and is proportional to the mass of the final state fermion.
The annihilation cross section is approximately given by
\begin{align}
\vev{\sigma v_{\text{rel}}}_0 &\simeq \vev{\sigma v_{\text{rel}} (\chi_1\chi_1\to b\bar{b},\tau\bar{\tau})}_0
\nonumber \\
&\simeq \frac{g_Z^2}{32\pi} (\lambda^Z)^2 \frac{3m_b^2+m_\tau^2}{m_Z^4} \quad (v_{\text{rel}}\to 0)
\nonumber \\
&\sim 2.8\times 10^{-26}\cdot (\lambda^Z)^2 \;\text{cm}^3/s
\nonumber \\
&\sim 1.8\times 10^{-30} \lrfp{300~\GeV}{|\mu|}{4} \left(1-\frac{m_{\chi_1}^2}{\mu^2}\right)^{-2} (\cos 2\beta)^2\;  \text{cm}^3/\text{s}\,,
\end{align}
where we have used running bottom quark mass $\overline{m}_b^{\overline{\text{MS}}}(100~\GeV)\simeq 3~\GeV$ (cf.~\cite{hep-ph/0503172}) in the third line, and Eq.~\eqref{eq:lambdaZapprox} in the last line.\footnote{
We omitted the correction around the region of $|m_{\chi_1}-m_Z/2|\lesssim \Gamma_Z$, where $\vev{\sigma v_{\text{rel}}}_0$ is further suppressed.
We have also checked by the micrOMEGAs, and the results agree within ${\cal O}(1\%)$ and ${\cal O}(10\%)$ for $\tau\bar{\tau}$ and $b\bar{b}$ modes, respectively.
}
Therefore, it is at most ${\cal O}(10^{-28})~\text{cm}^3\text{s}^{-1}$, and much smaller in most of the parameter space, which is smaller than the constraints such as the Fermi-LAT bounds in Ref.~\cite{Ackermann:2015zua}.

\item neutrinos from DM annihilation in the Sun.

Pair annihilations of DMs which are captured in the Sun generate neutrinos, and there have been searches for such neutrinos.
In the present scenario, the DM annihilation rate in the Sun is proportional to the following effective SD scattering cross section~\cite{hep-ph/0404175}
\begin{align}
\sigma_p^{\text{SD(eff)}} = \sigma_{p}^{\text{SD}} \tanh^2 
\left( \sqrt{\Gamma_{\text{cap.}} \Gamma_{\text{ann.}}}\cdot t_\odot \right)\,,
\end{align}
where $\sigma_p^{\text{SD}}$ is given in Eq.~\eqref{eq:SDXS}, $\Gamma_{\text{cap.}}$ and $\Gamma_{\text{ann.}}$ are the capture and annihilation rates of DM in the Sun, respectively, and $t_\odot$ is the age of the solar system. 
If $\sqrt{\Gamma_{\text{cap.}} \Gamma_{\text{ann.}}}\cdot t_\odot\gg 1$, the capture and annihilation rates are in equilibrium. 
In the present case, it is given by~\cite{hep-ph/0404175}
\begin{align}
\sqrt{\Gamma_{\text{cap.}} \Gamma_{\text{ann.}}}\cdot t_\odot
\simeq
1.3\times \lrfp{\sigma_p^{\text{SD}}}{10^{-40}\text{cm}^2}{1/2} \lrfp{\vev{\sigma v_{\text{rel}}}_0}{10^{-29}\text{cm}^3/\text{s}}{1/2} \lrfp{50~\text{GeV}}{m_{\chi_1}}{1/4}\,,
\end{align}
and hence the annihilation rate is not completely saturated by the scattering rate.
Currently the Super-Kamiokande gives the strongest bound~\cite{1503.04858} in the mass range of our interest, which are given by $\sigma_p^{\text{SD(eff)}}\cdot \text{Br}(\tau\bar{\tau})\lesssim (1 \text{--} 2)\times 10^{-40}\text{cm}^2$ and $\sigma_p^{\text{SD(eff)}}\cdot \text{Br}(b\bar{b})\lesssim (2 \text{--} 3)\times 10^{-39}\text{cm}^2$ for $m_{\chi_1}\simeq (30 \text{--} 70)~\GeV$.
Here, $\text{Br}(X)=\vev{\sigma v_{\text{rel}}(\chi\chi\to X)}_0/\vev{\sigma v_{\text{rel}}(\chi\chi\to \text{all})}_0$ represent the branching fractions of the annihilation channels, which are given by $\text{Br}(\tau\bar{\tau})\sim m_\tau^2/(3m_b^2+m_\tau^2)$ and $\text{Br}(b\bar{b})\sim 3m_b^2/(3m_b^2+m_\tau^2)$ in the present scenario.
As a result, we found that the bound from the neutrinos are weaker than the bound on $\sigma^{\text{SD}}$ from the XENON~100~\cite{1301.6620} for $m_{\chi_1}\simeq (30 \text{--} 70)~\GeV$ (see Sec.~\ref{subsec:direct}).

\item
mono-jet and mono-photon.
\\
Mono-photon events could be produced at the LEP, via $e^+e^-\to \gamma Z^*\to \gamma \chi_1\chi_1$, but it is expected that the constraint is very weak (see, e.g., Ref.~\cite{hep-ex/0406019}).
The mono-jet events at the LHC, $pp\to j Z^*\to j \chi_1\chi_1$, also gives only a weak constraint on $\lambda^Z$ compared to the other bounds. (See, e.g., a study on generic vector mediator in Ref.~\cite{1407.8257}.)

\end{itemize}

\section{Summary}
\label{sec:discussion}
In this paper, we have investigated the Higgs- and $Z$-resonant neutralino DM scenario.
The phenomenology of this scenario is determined only by three parameters, Bino mass $M_1$, Higgsino mass $\mu$ and $\tan\beta$ when all other SUSY particles and heavy Higgs bosons are decoupled.
In this scenario, the Bino-like neutralino DM can have the correct thermal relic abundance via the Higgs- and $Z$-resonant annihilations. 
We have investigated the current constraints and future prospects comprehensively for essentially all the parameter space. 

As constraints, we have included: (i) relic abundance $\Omega_\chi h^2 = 0.120$~\cite{Agashe:2014kda,1502.01589}, (ii) direct detection constraints on the spin-independent (SI) scattering cross section from the LUX~\cite{1310.8214} and on the spin-dependent (SD) scattering cross section from the XENON 100~\cite{1301.6620},
(iii)  constraint on the Higgs invisible decay from global fit~\cite{1303.3570,1306.2941}.
For future prospects, we have investigated: (i) prospects of the XENON 1T for the SI~\cite{1206.6288} and SD~\cite{1211.4573} scattering cross sections of DM direct detection, (ii) expected sensitivity of the HL-LHC~\cite{1309.7925,1310.8361} and of the ILC~\cite{1310.0763} for the Higgs invisible decay, and (iii) expected sensitivity of the LHC chargino/neutralino search at 14 TeV for 300 fb$^{-1}$ and 3000 fb$^{-1}$~\cite{ATL-PHYS-PUB-2014-010}.

\begin{figure}[t]
\begin{center}
\includegraphics[width=6.5cm]{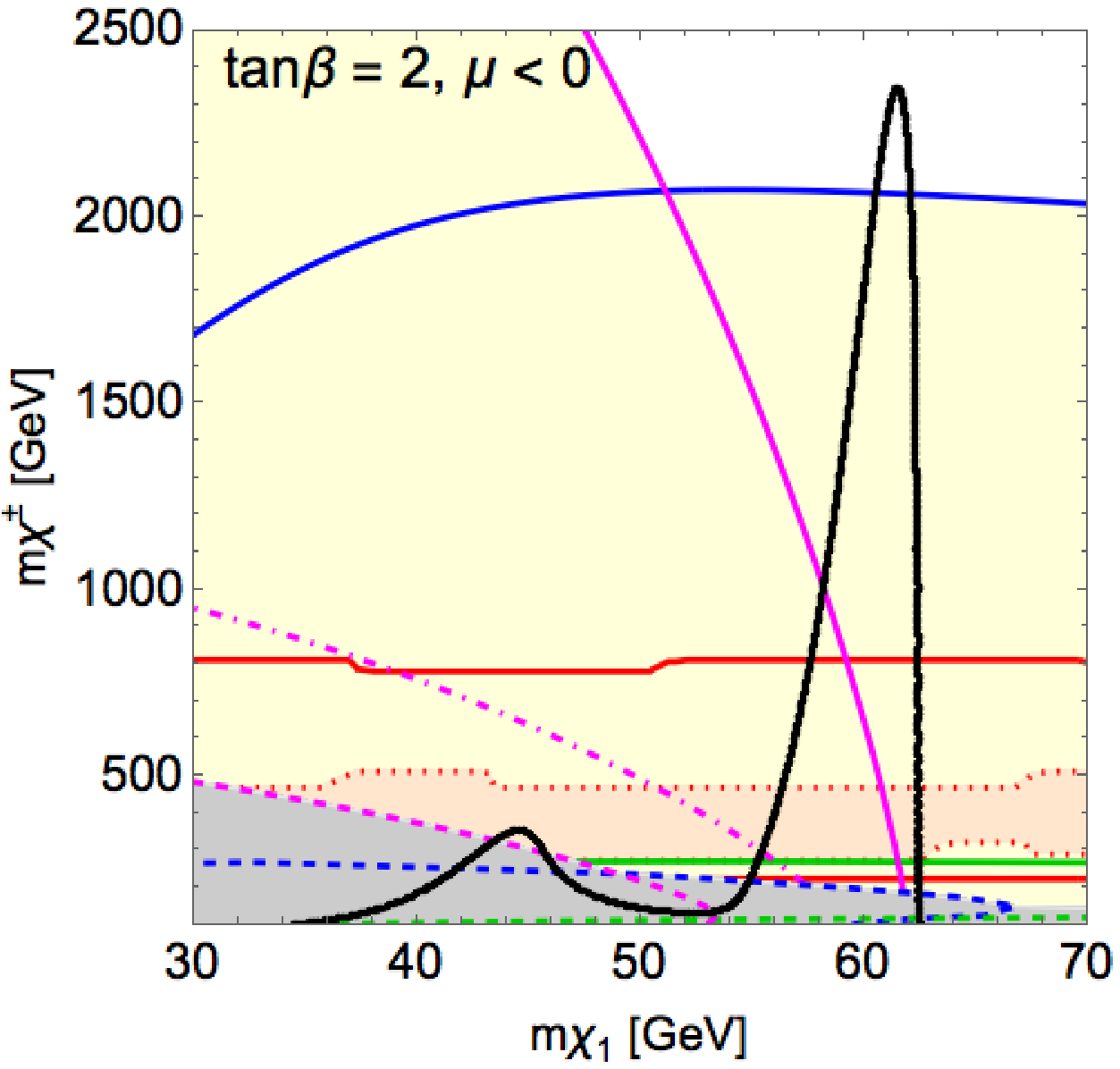}
\includegraphics[width=6.5cm]{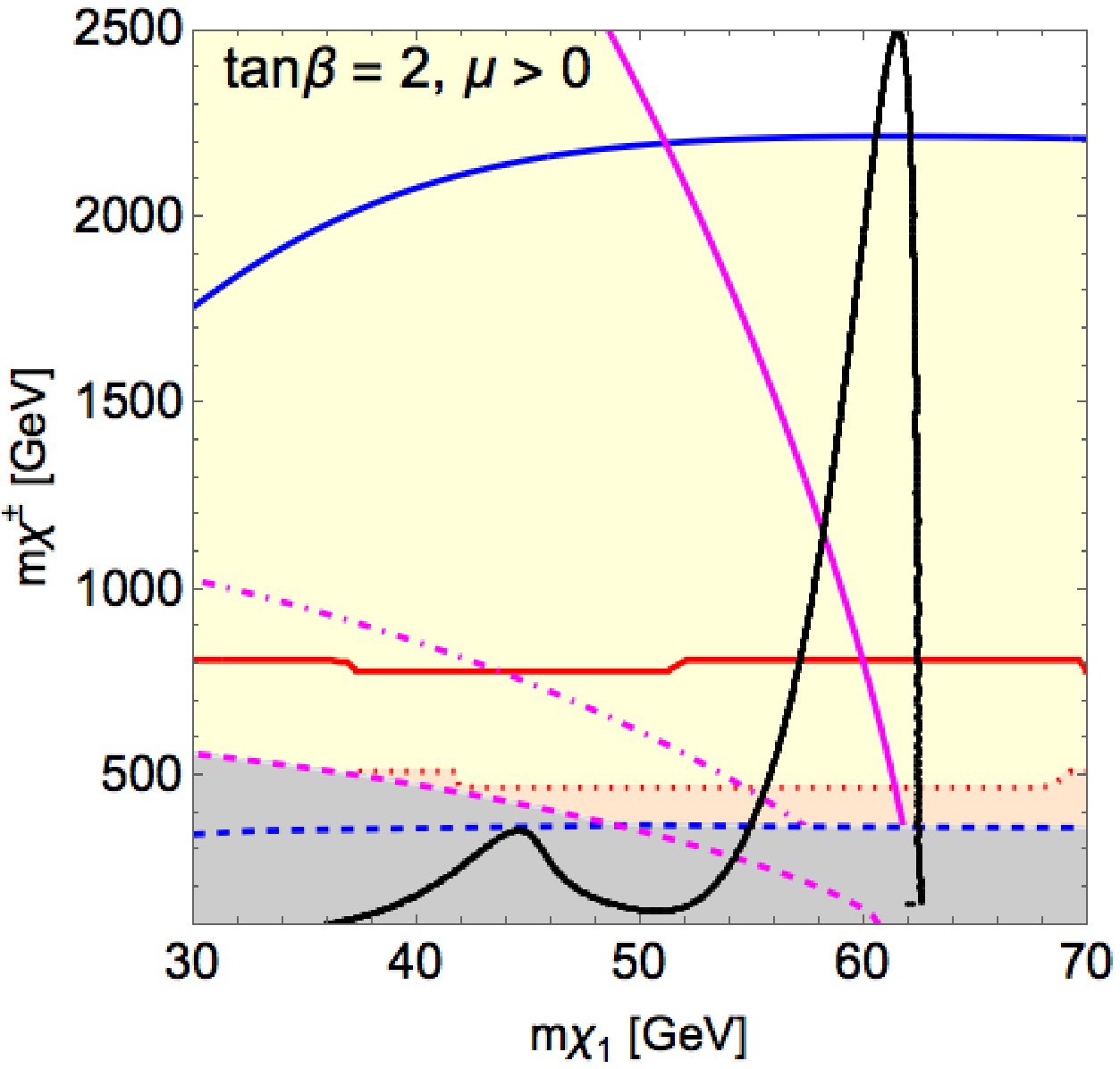}
\\
\includegraphics[width=6.5cm]{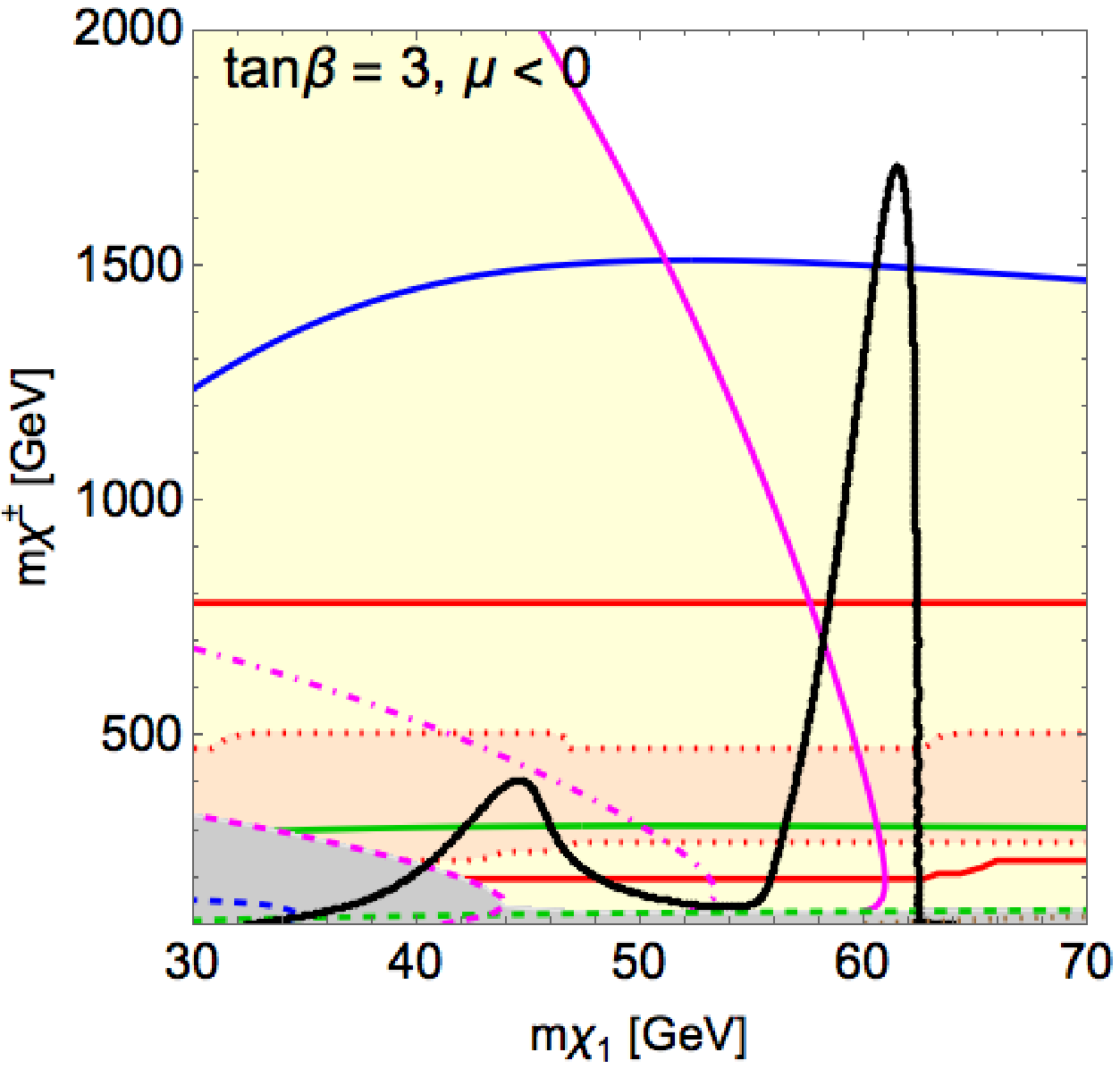}
\includegraphics[width=6.5cm]{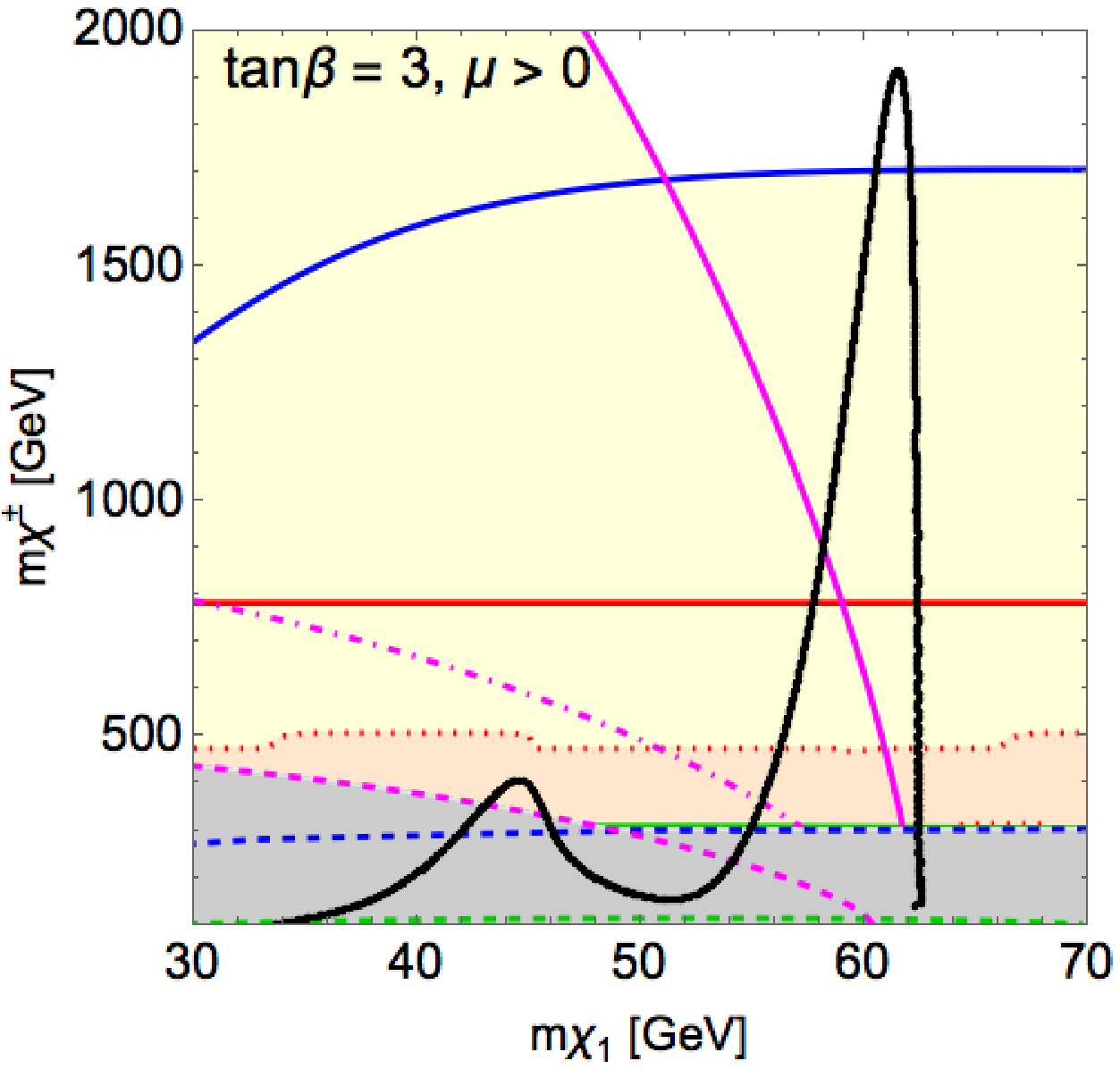}
\caption{
Constraints and future sensitivity of the Higgs- and $Z$-resonant neutralino DM, for $\tan\beta = 2$ and 3, and for $\mu<0$ (left) and $\mu>0$ (right).
The black lines show abundance $\Omega_{\chi_1}\simeq 0.120$.
The gray shaded region is excluded by current constraints;
the LUX bound on $\sigma^{\text{SI}}_N$ (blue dashed), the XENON100 bound on $\sigma^{\text{SD}}_n$ (green dashed), and the Higgs invisible decay (magenta dashed).
The light yellow region will be probed by future experiments;
the XENON 1T via SI-scattering (blue solid) and SD-scattering (green solid), the Higgs invisible decay at the HL-LHC (magenta dot-dashed) and at the ILC (magenta solid), and the search for the chargino and neutralinos at the 14 TeV LHC, at $3000~\text{fb}^{-1}$ (red solid).
In particular, the light orange region within the red dotted lines will be reached at $300~\text{fb}^{-1}$.
See text for details.
}
\label{fig:main_tb2and3}
\end{center}
\end{figure}

\begin{figure}[t]
\begin{center}
\includegraphics[width=6.5cm]{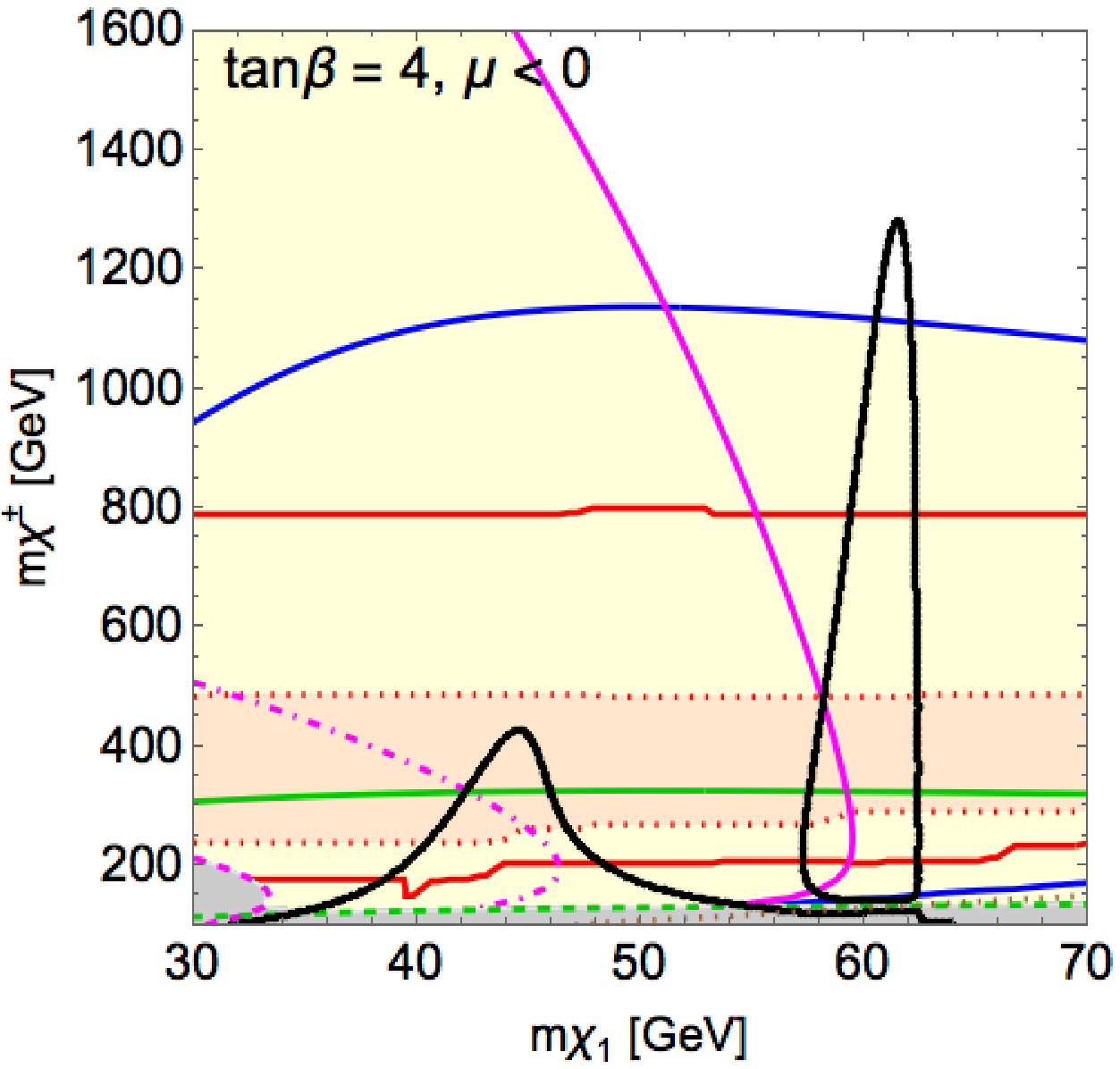}
\includegraphics[width=6.5cm]{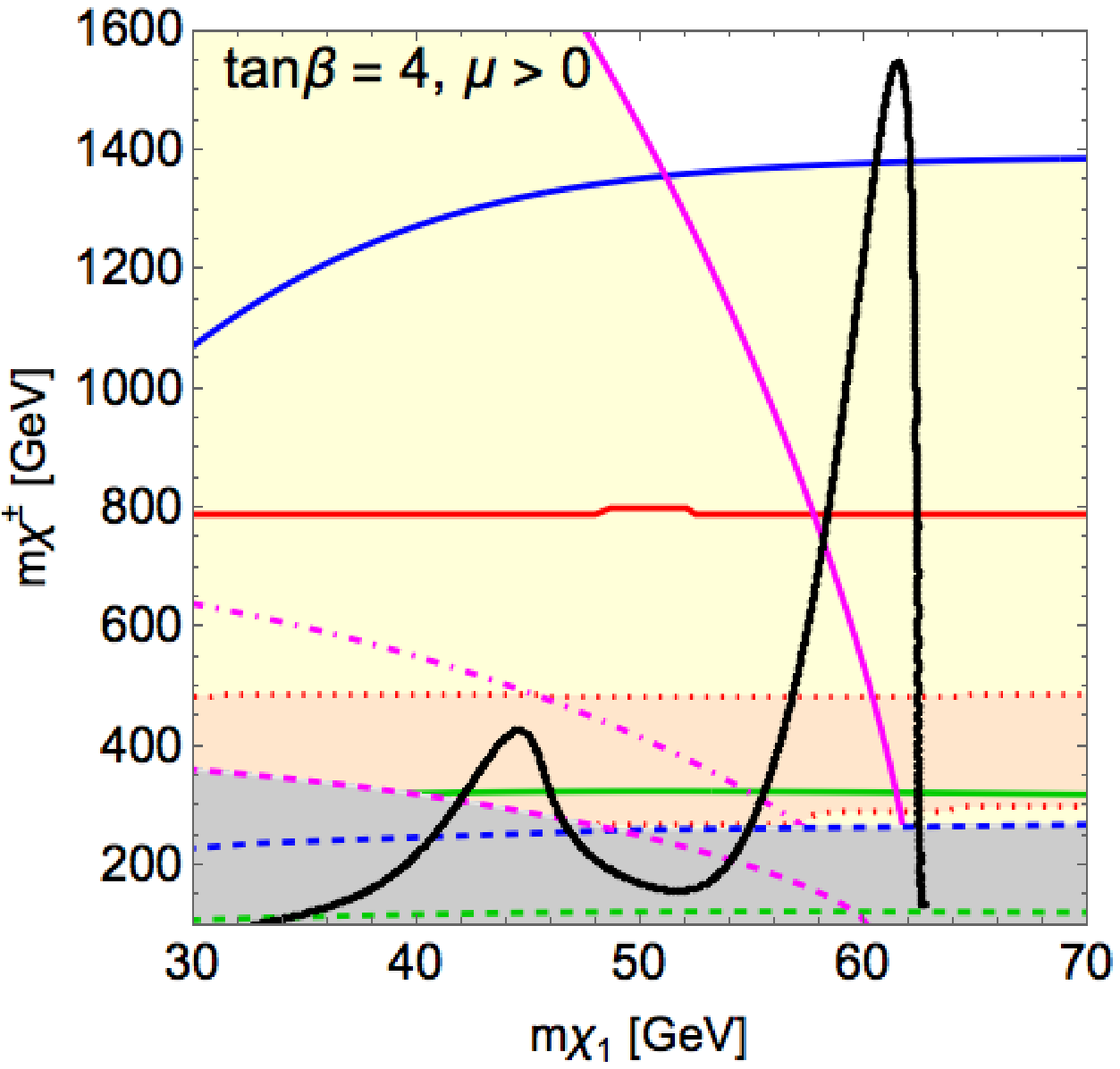}
\\
\includegraphics[width=6.5cm]{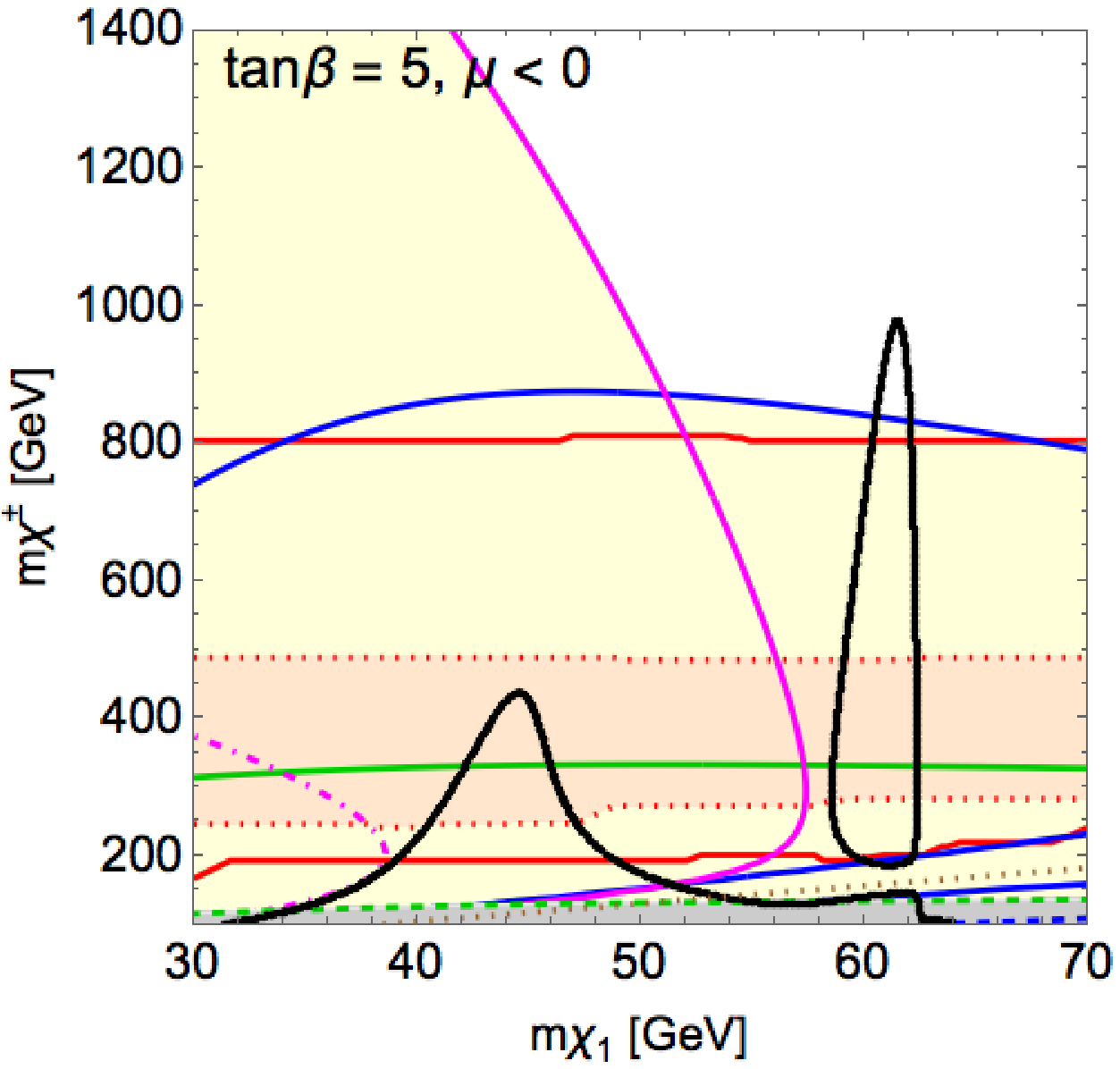}
\includegraphics[width=6.5cm]{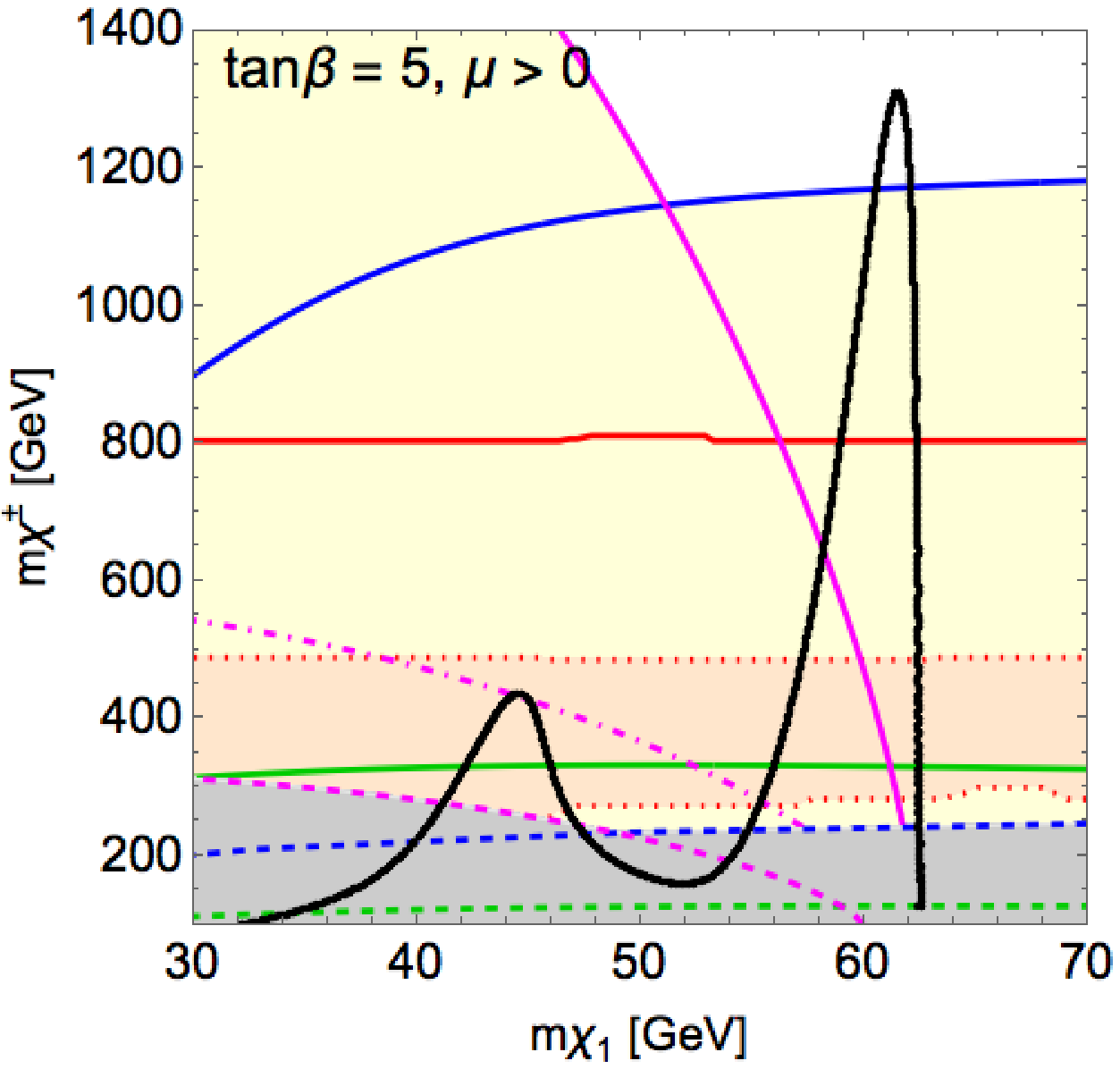}
\\
\includegraphics[width=6.5cm]{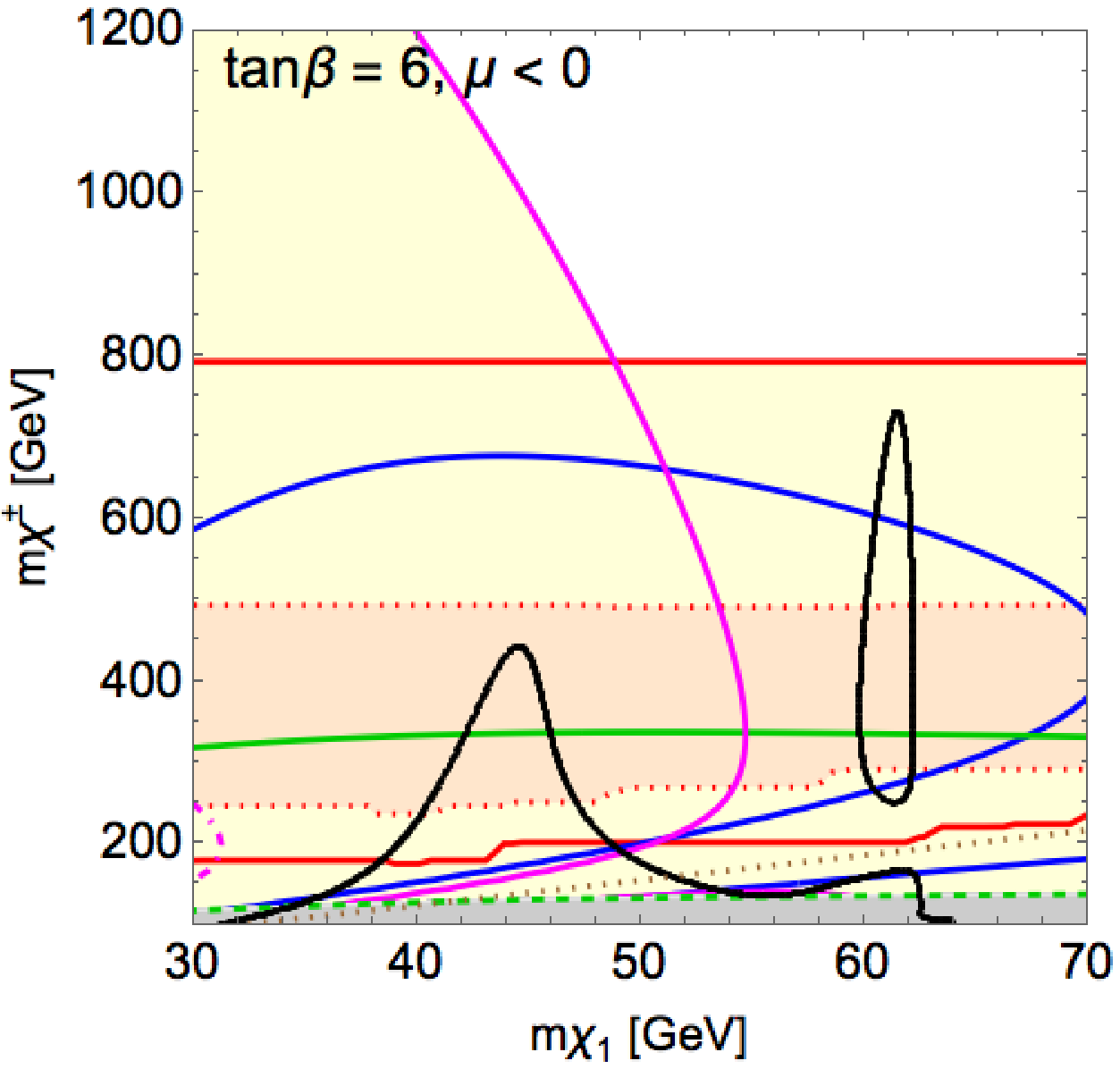}
\includegraphics[width=6.5cm]{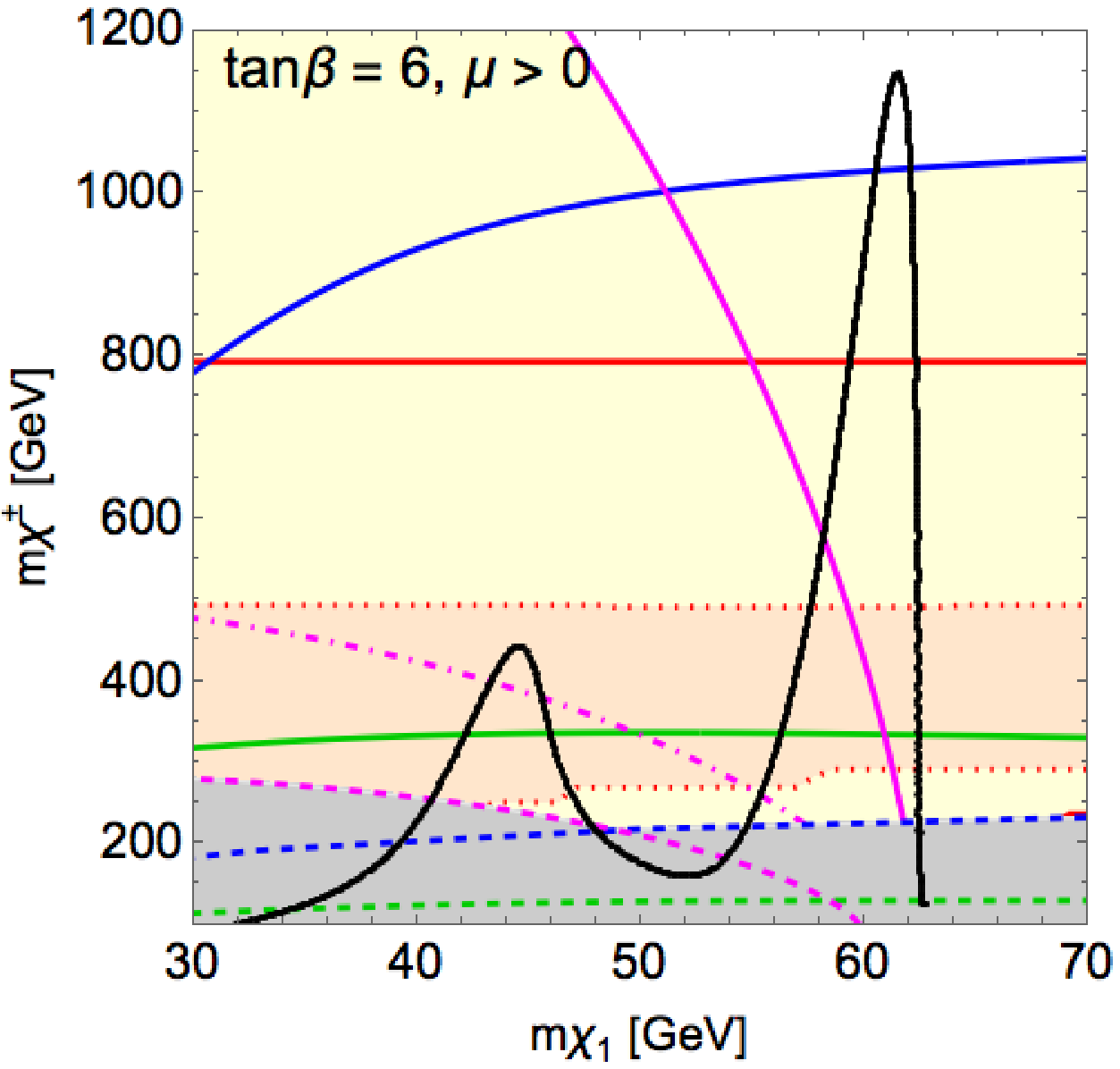}
\caption{The same as Fig.~\ref{fig:main_tb2and3}, but for $\tan\beta=4$, 5 and 6.
For $\mu<0$, the blind spot $\lambda^h=0$ is shown with a brown dotted line.}
\label{fig:main_tb45and6}
\end{center}
\end{figure}

\begin{figure}[t]
\begin{center}
\includegraphics[width=6.5cm]{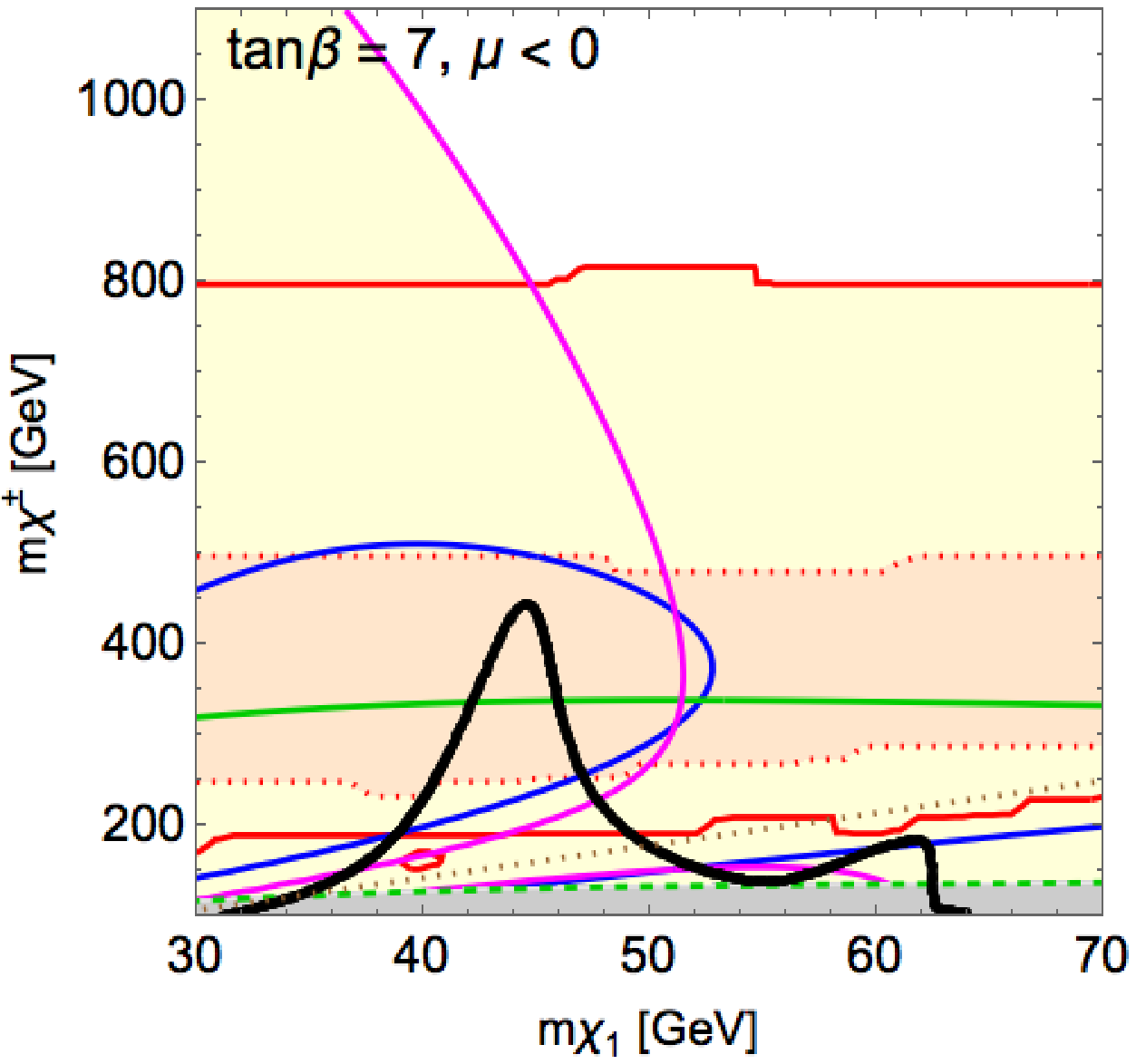}
\includegraphics[width=6.5cm]{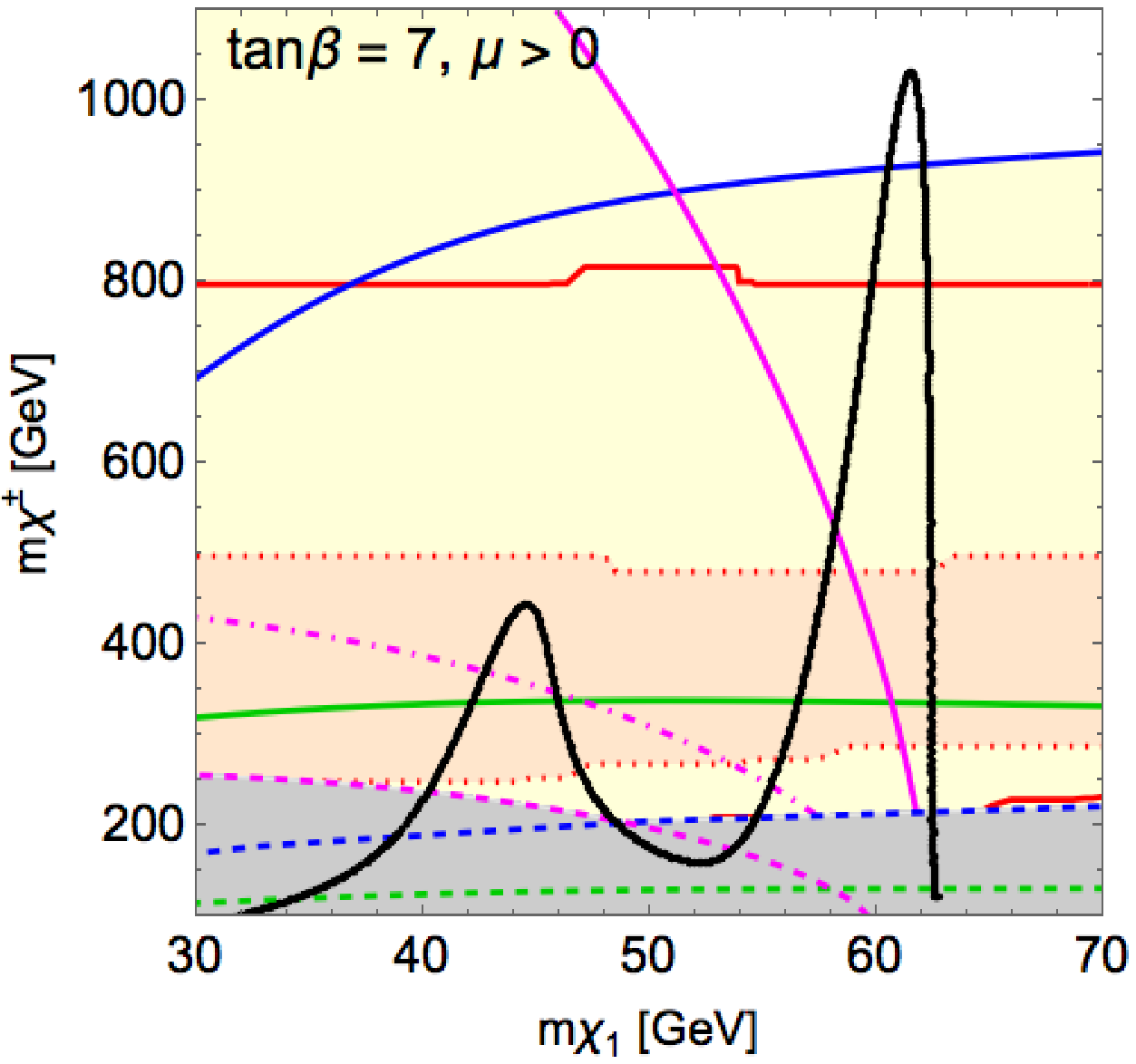}
\\
\includegraphics[width=6.5cm]{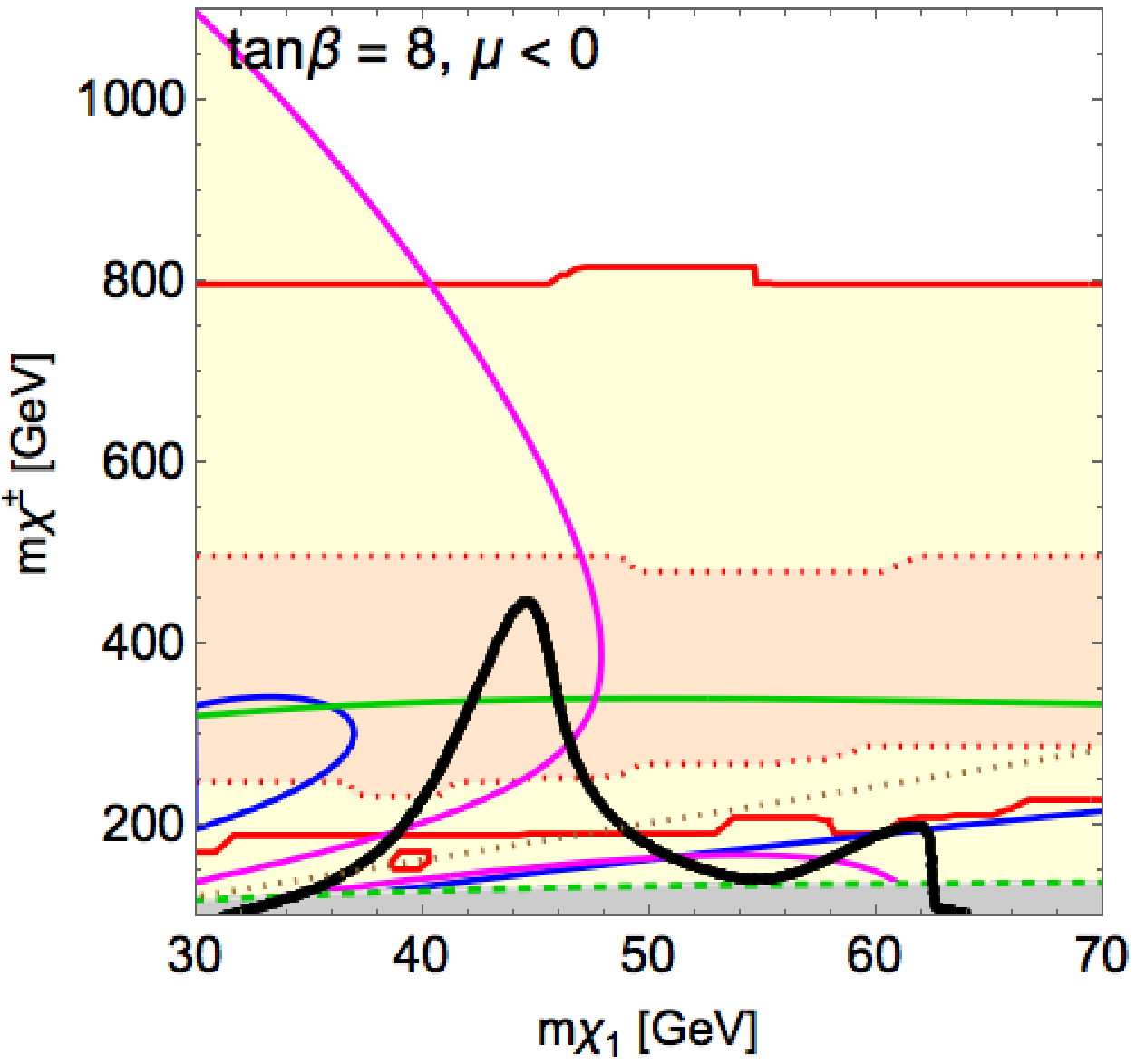}
\includegraphics[width=6.5cm]{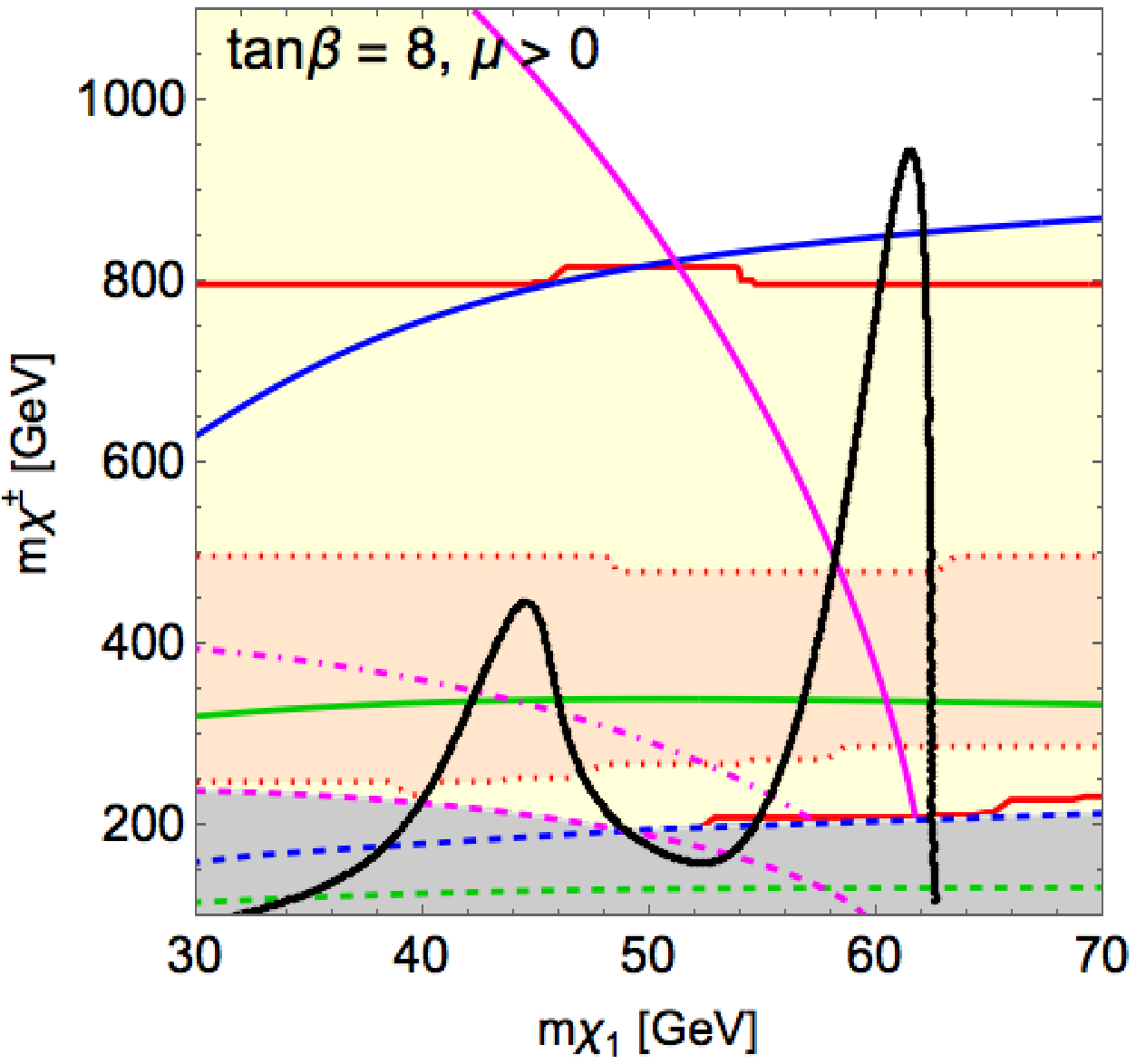}
\\
\includegraphics[width=6.5cm]{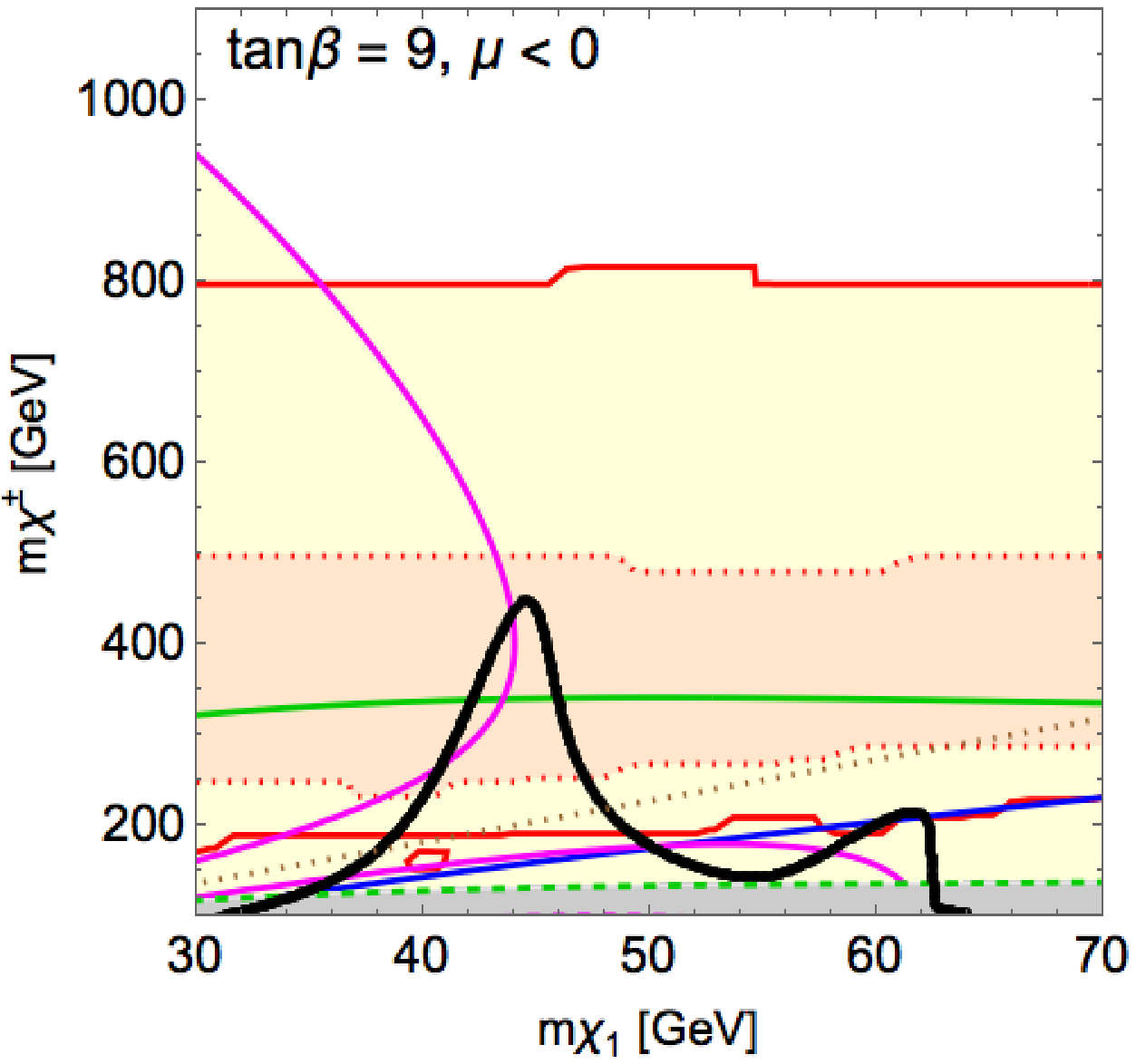}
\includegraphics[width=6.5cm]{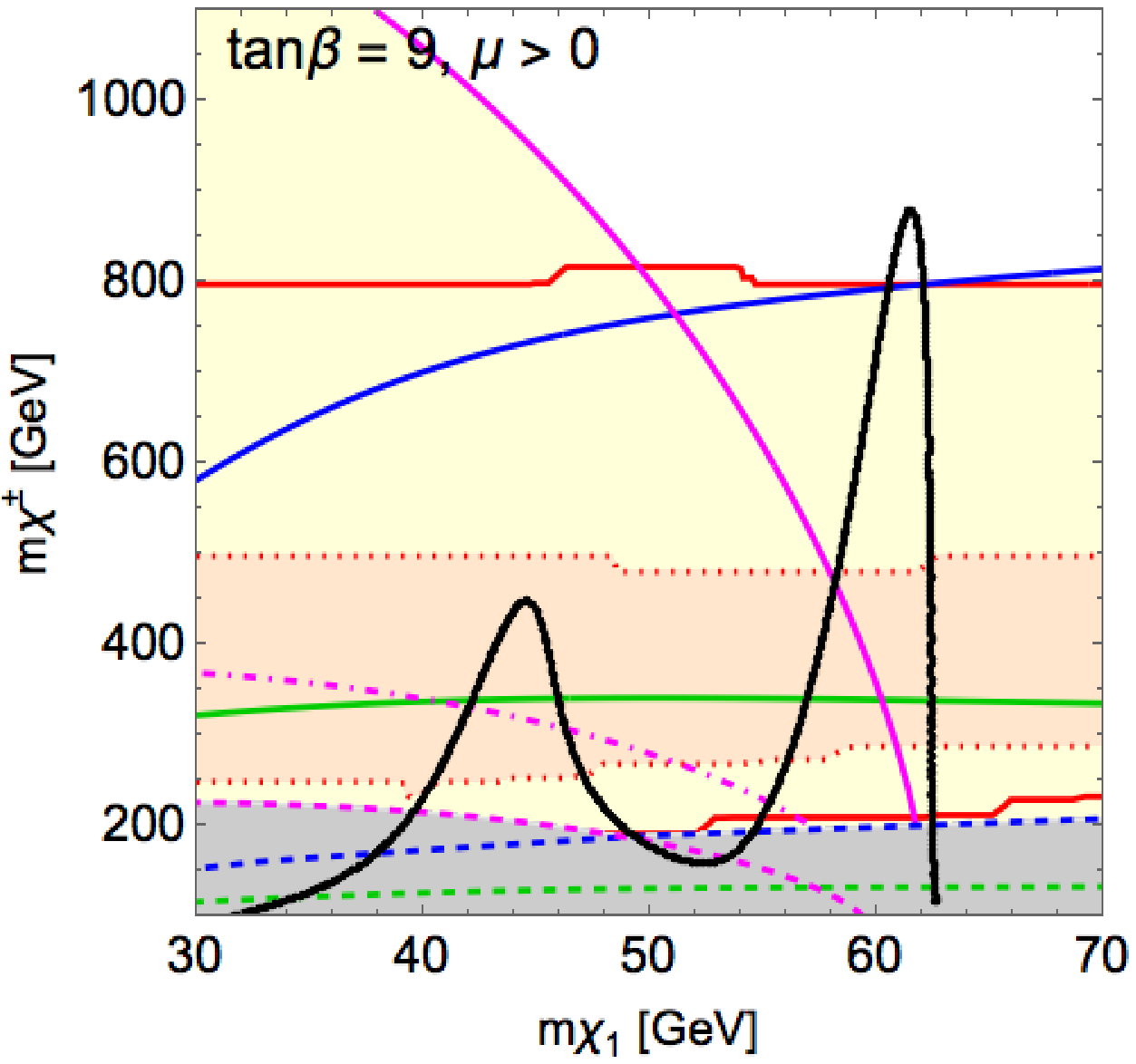}
\caption{The same as Fig.~\ref{fig:main_tb2and3}, but for $\tan\beta=7$, 8 and 9.}
\label{fig:main_tb78and9}
\end{center}
\end{figure}

\begin{figure}[t]
\begin{center}
\includegraphics[width=6.5cm]{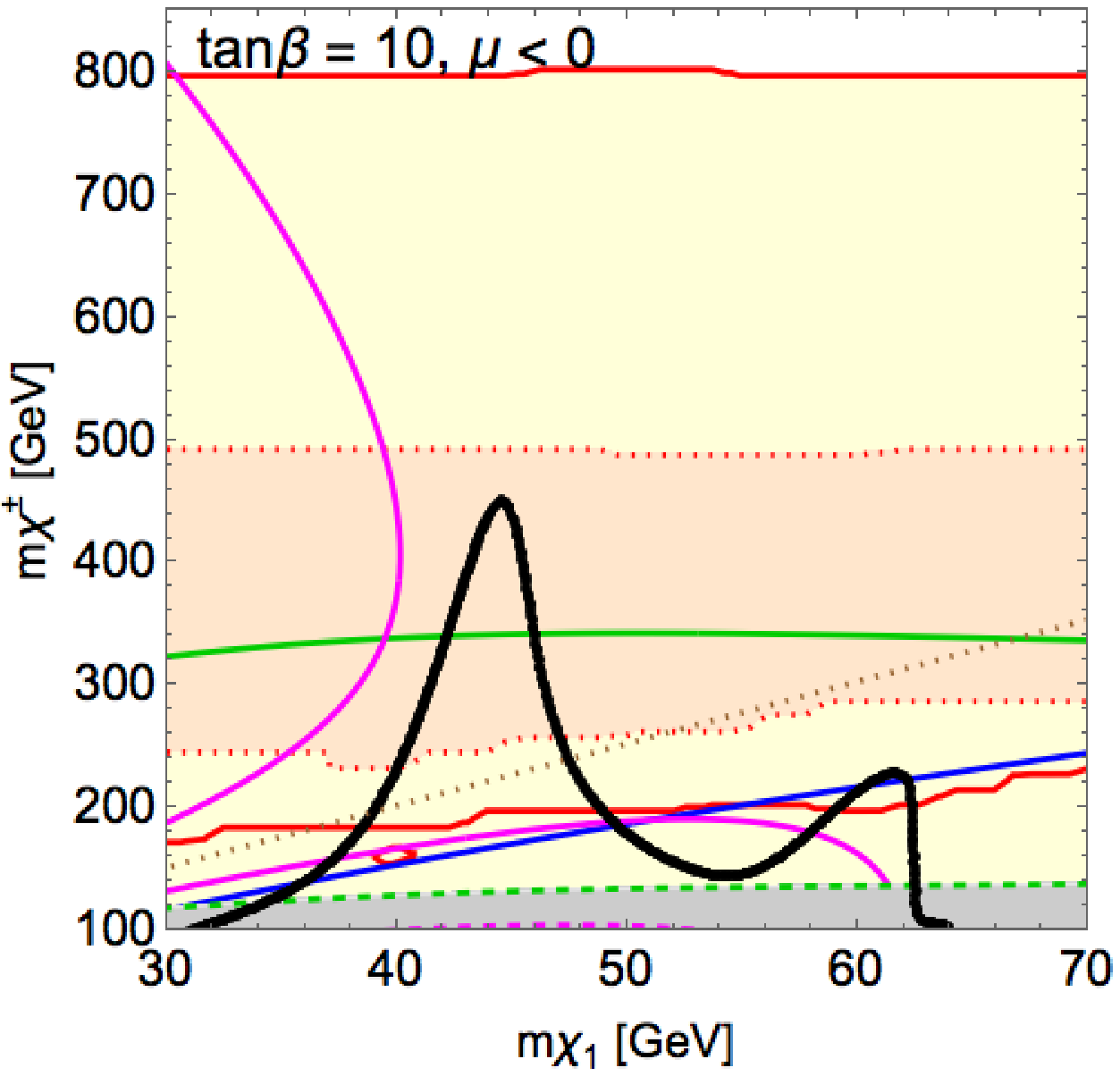}
\includegraphics[width=6.5cm]{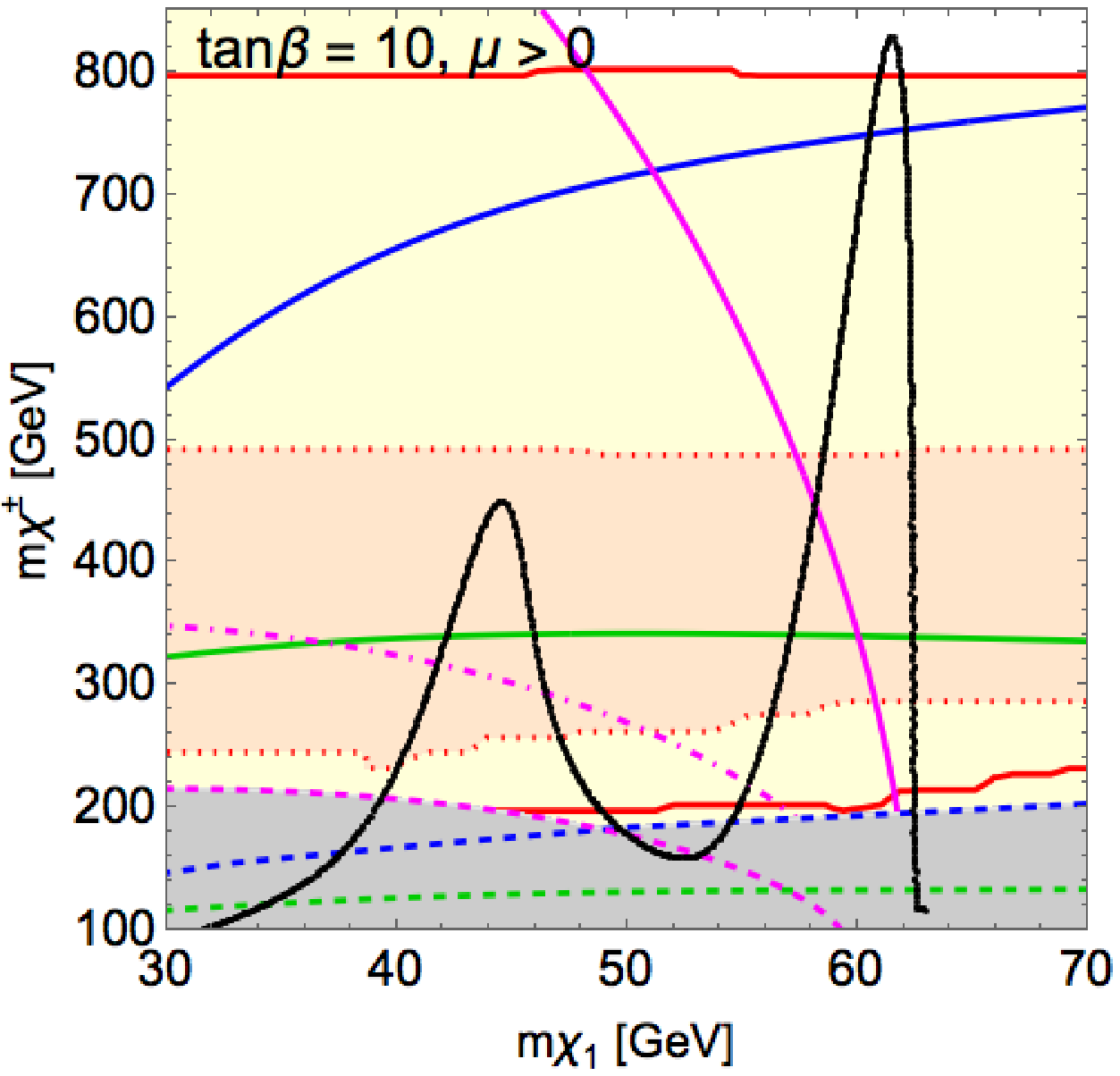}
\\
\includegraphics[width=6.5cm]{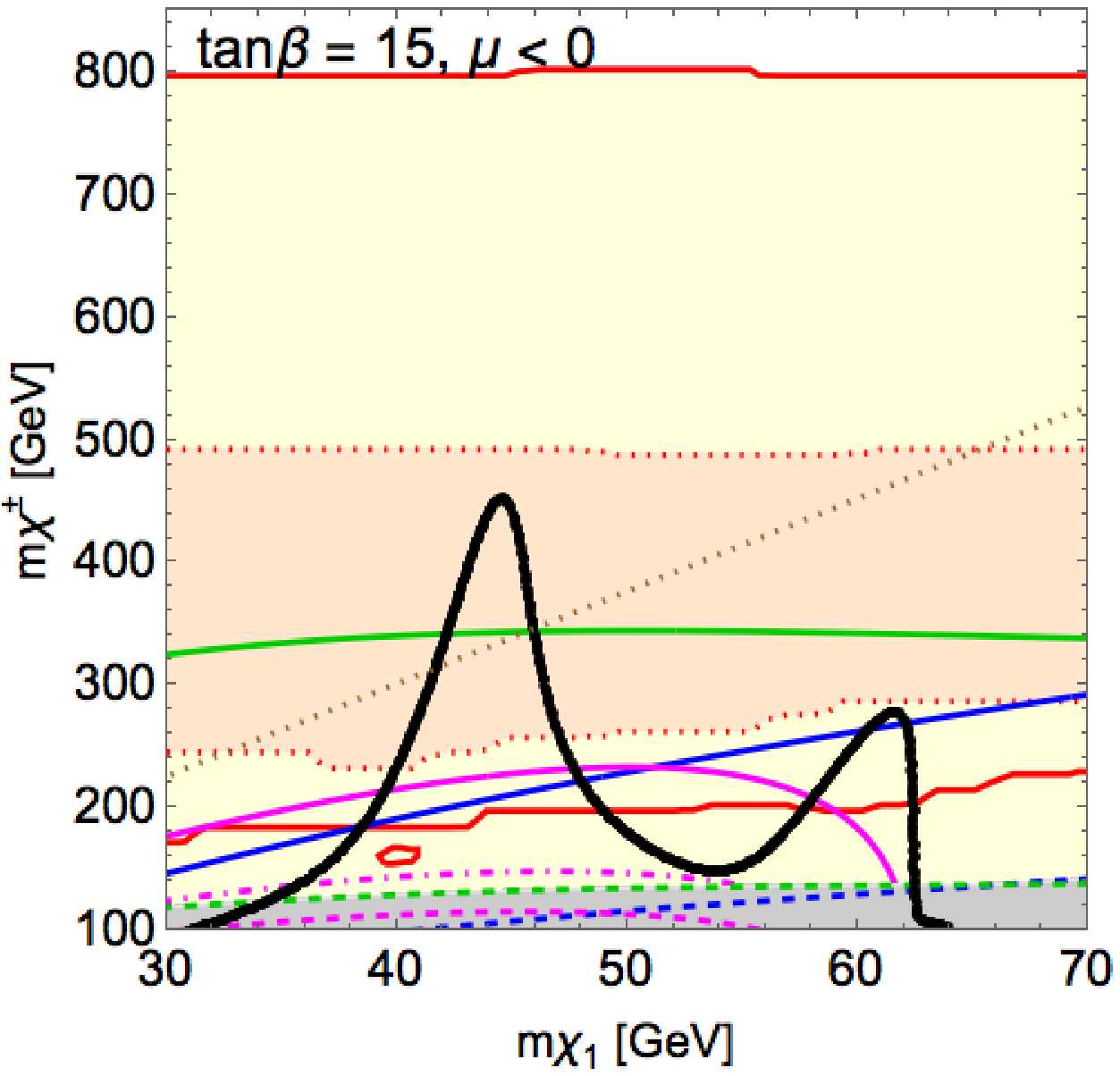}
\includegraphics[width=6.5cm]{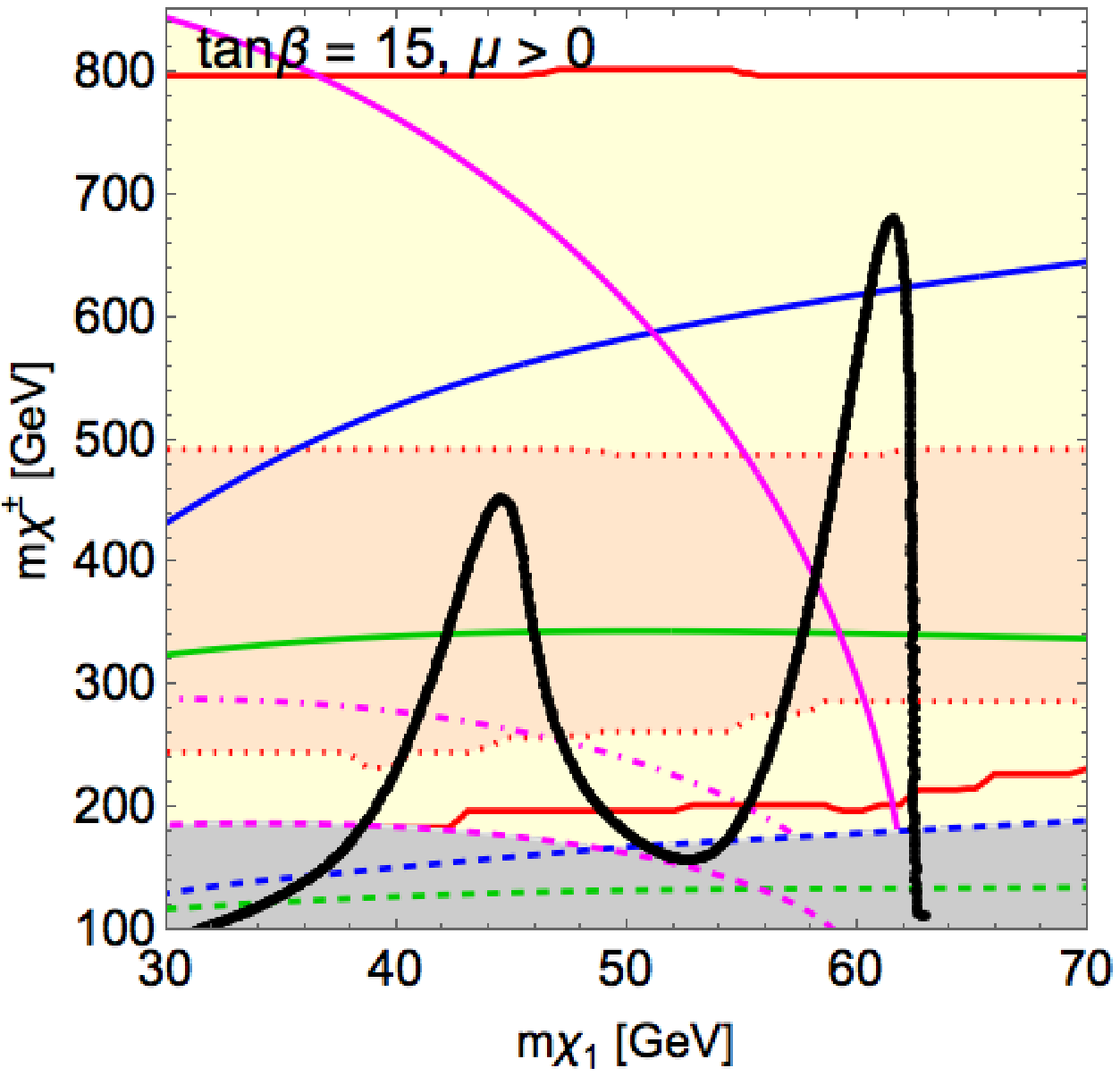}
\\
\includegraphics[width=6.5cm]{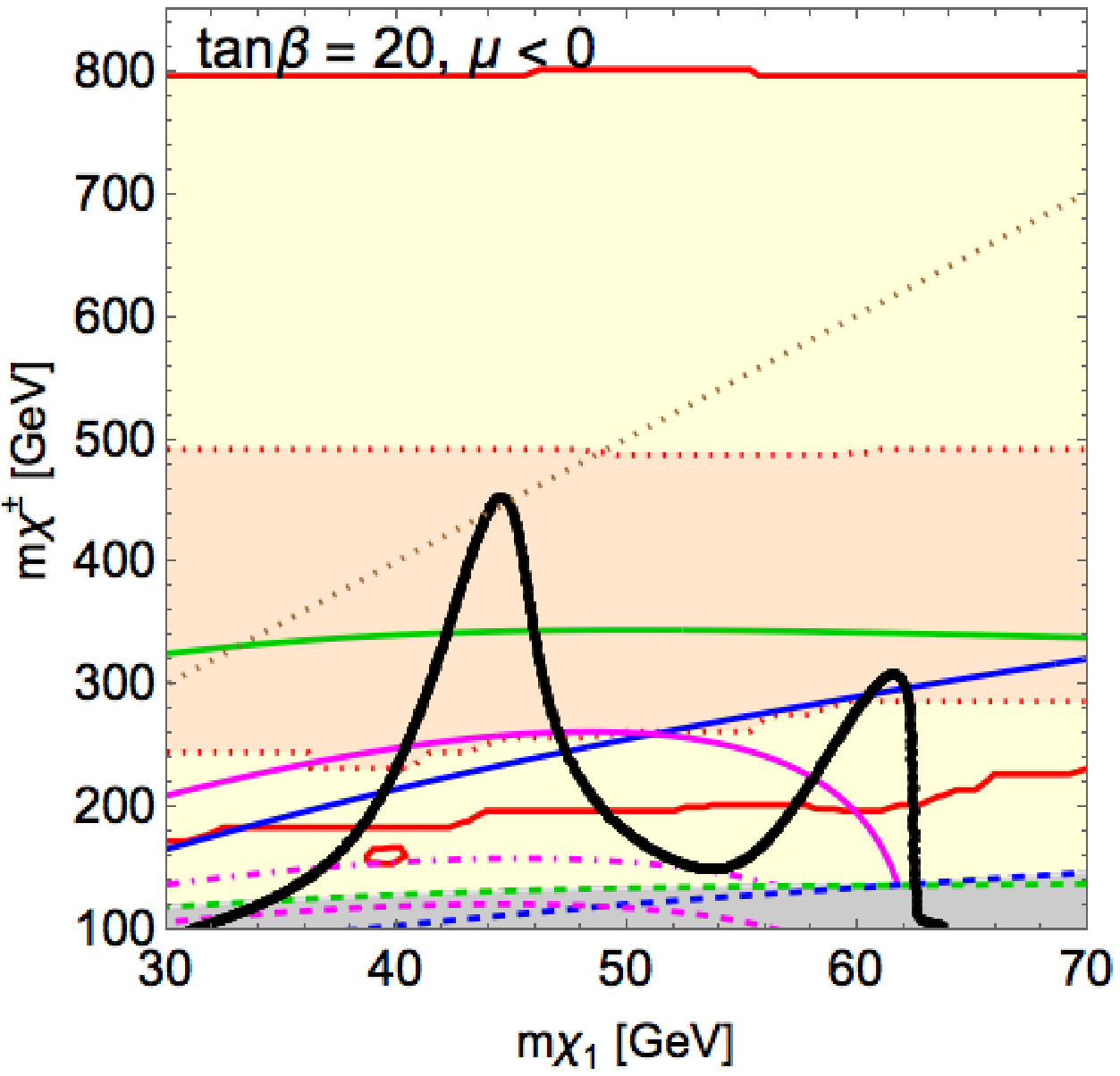}
\includegraphics[width=6.5cm]{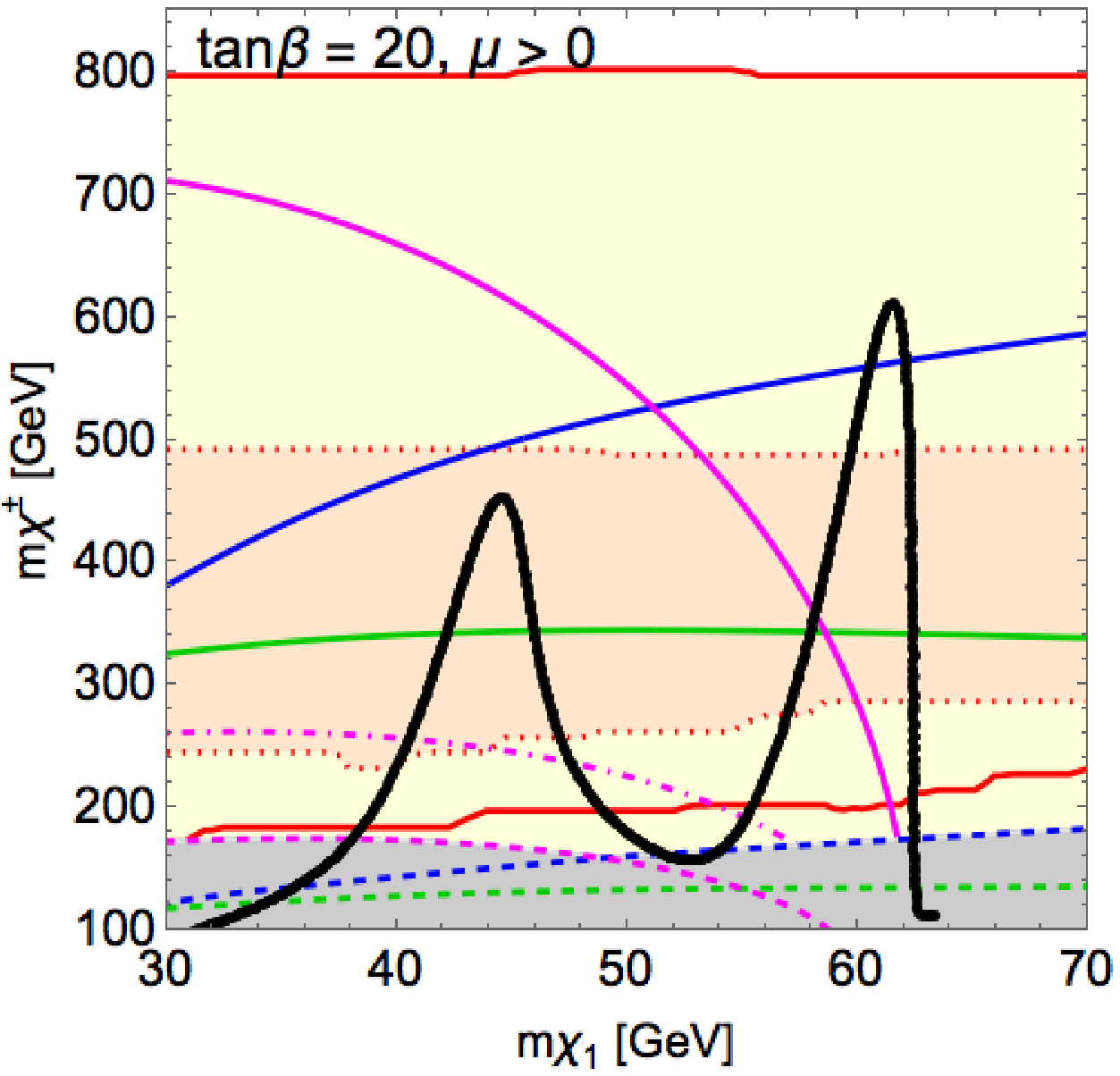}
\caption{The same as Fig.~\ref{fig:main_tb2and3}, but for $\tan\beta=10$, 15 and 20.}
\label{fig:main_tb1015and20}
\end{center}
\end{figure}

\begin{figure}[t]
\begin{center}
\includegraphics[width=6.5cm]{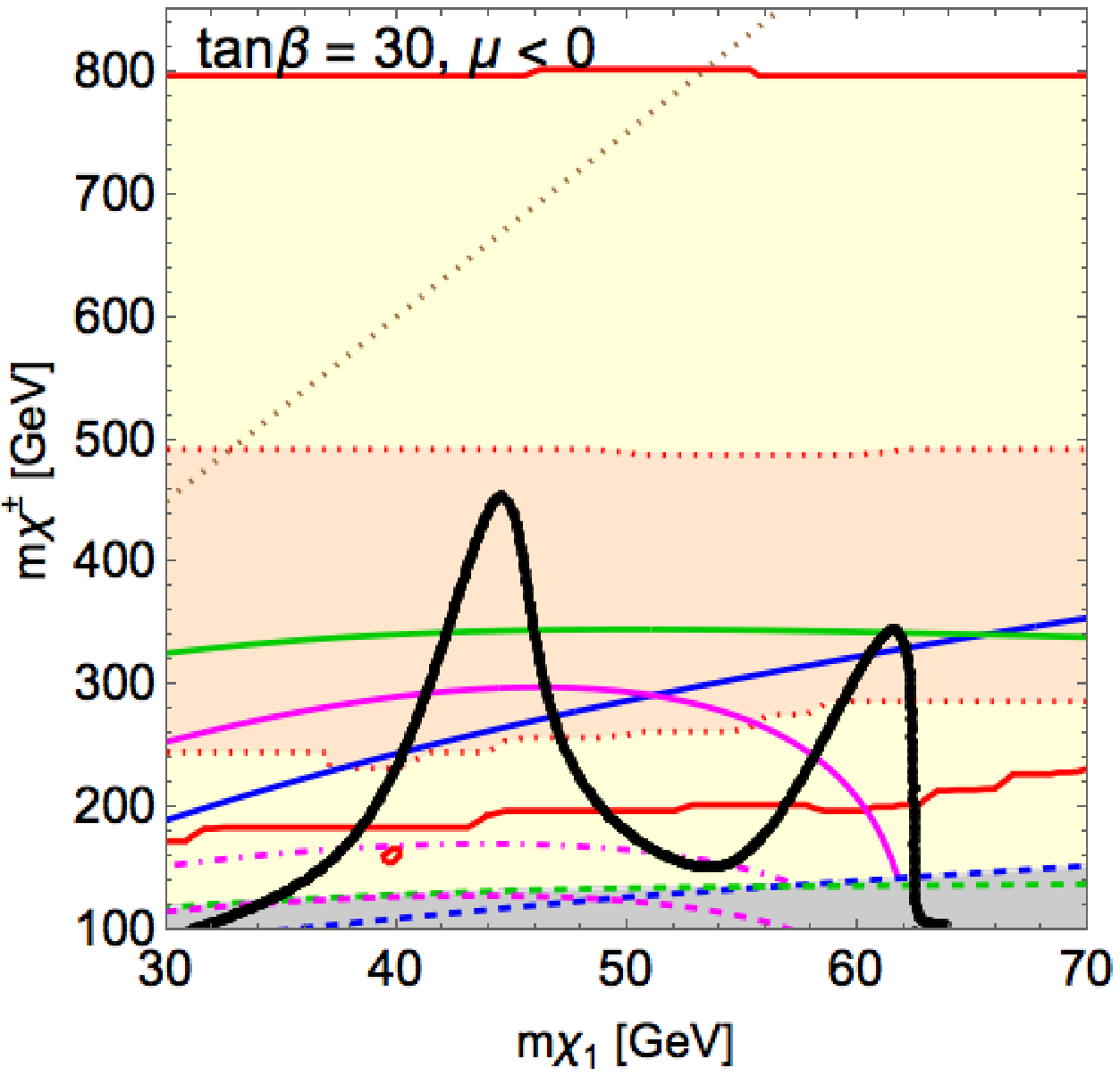}
\includegraphics[width=6.5cm]{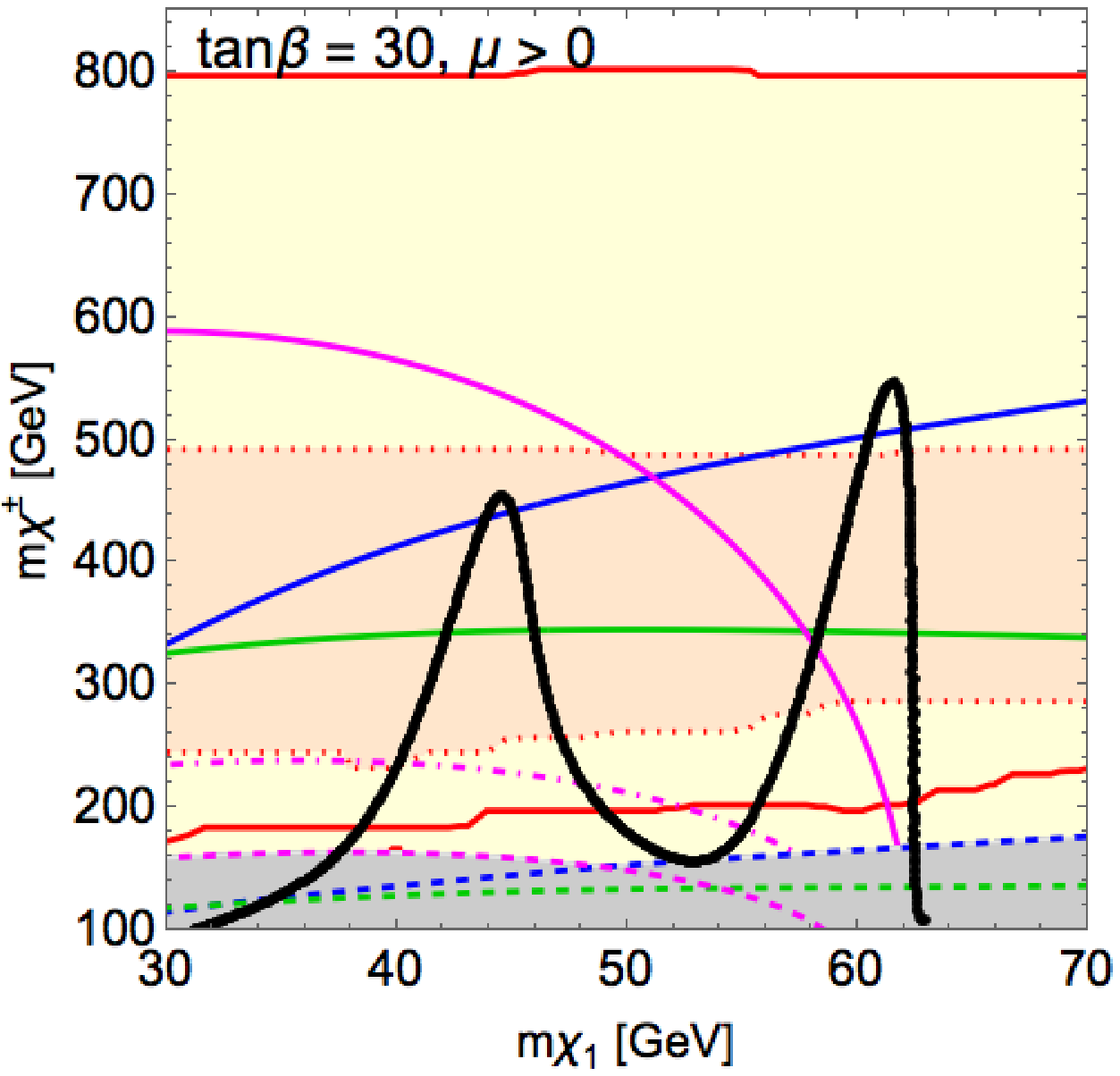}
\\
\includegraphics[width=6.5cm]{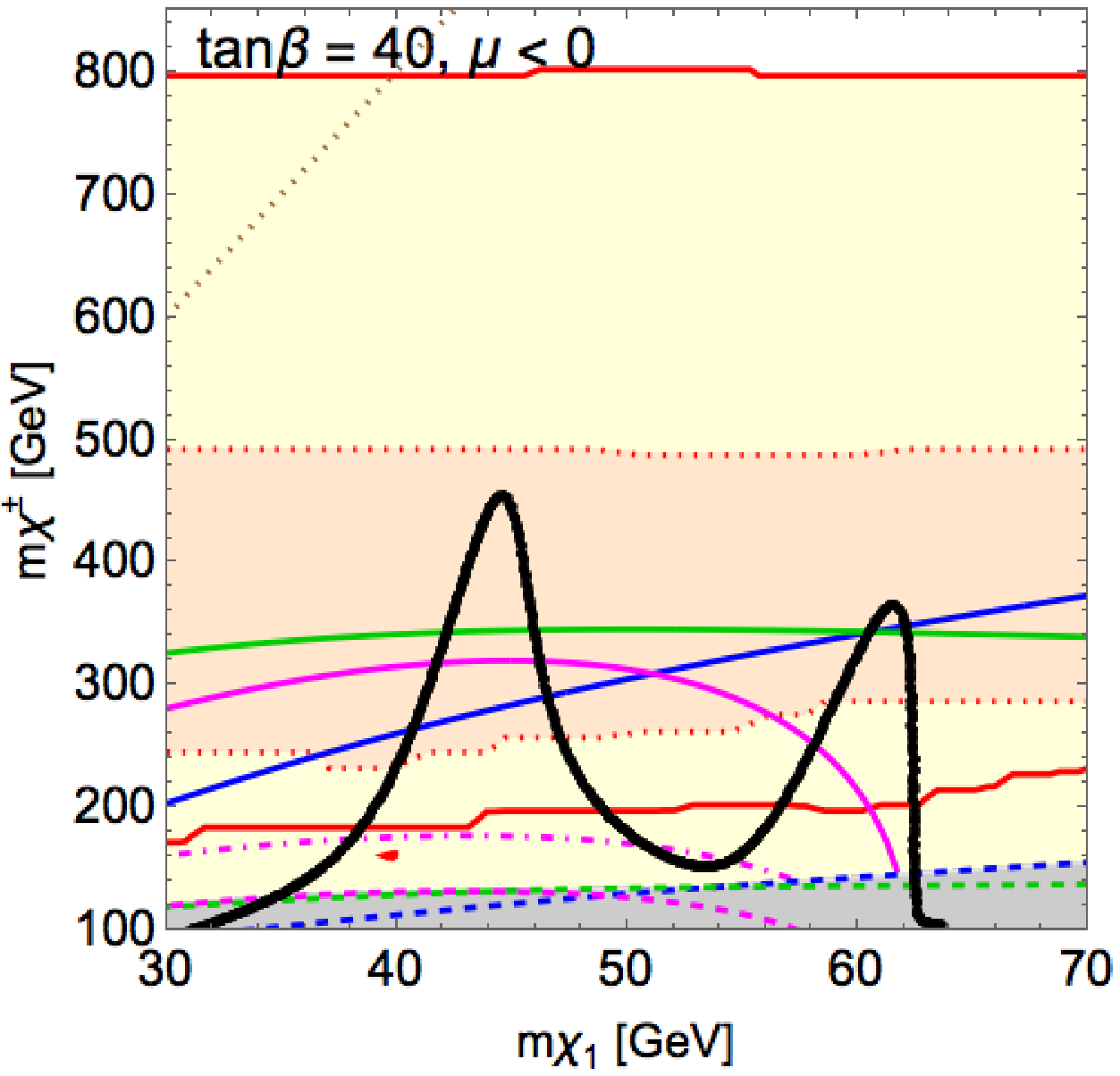}
\includegraphics[width=6.5cm]{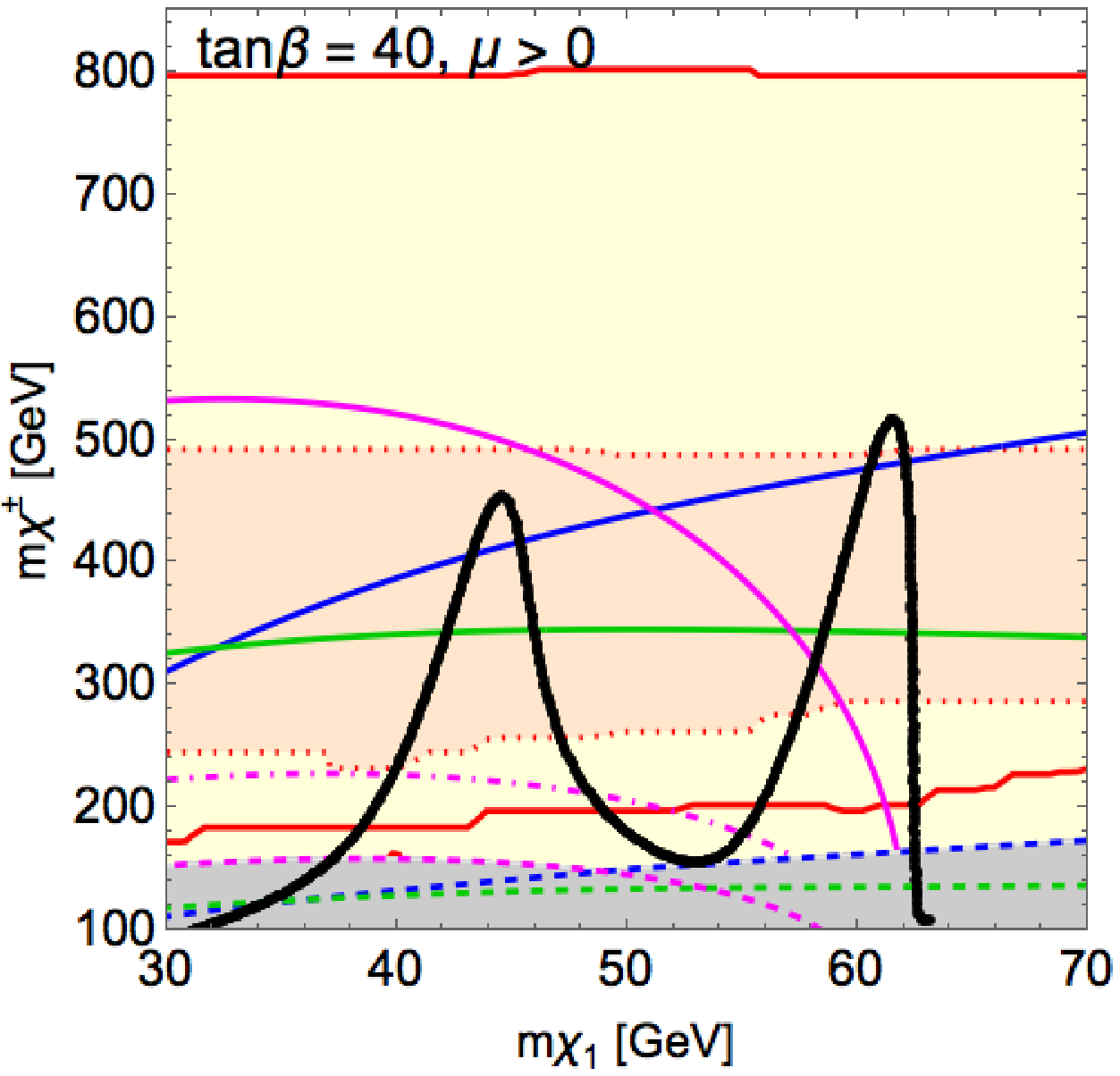}
\\
\includegraphics[width=6.5cm]{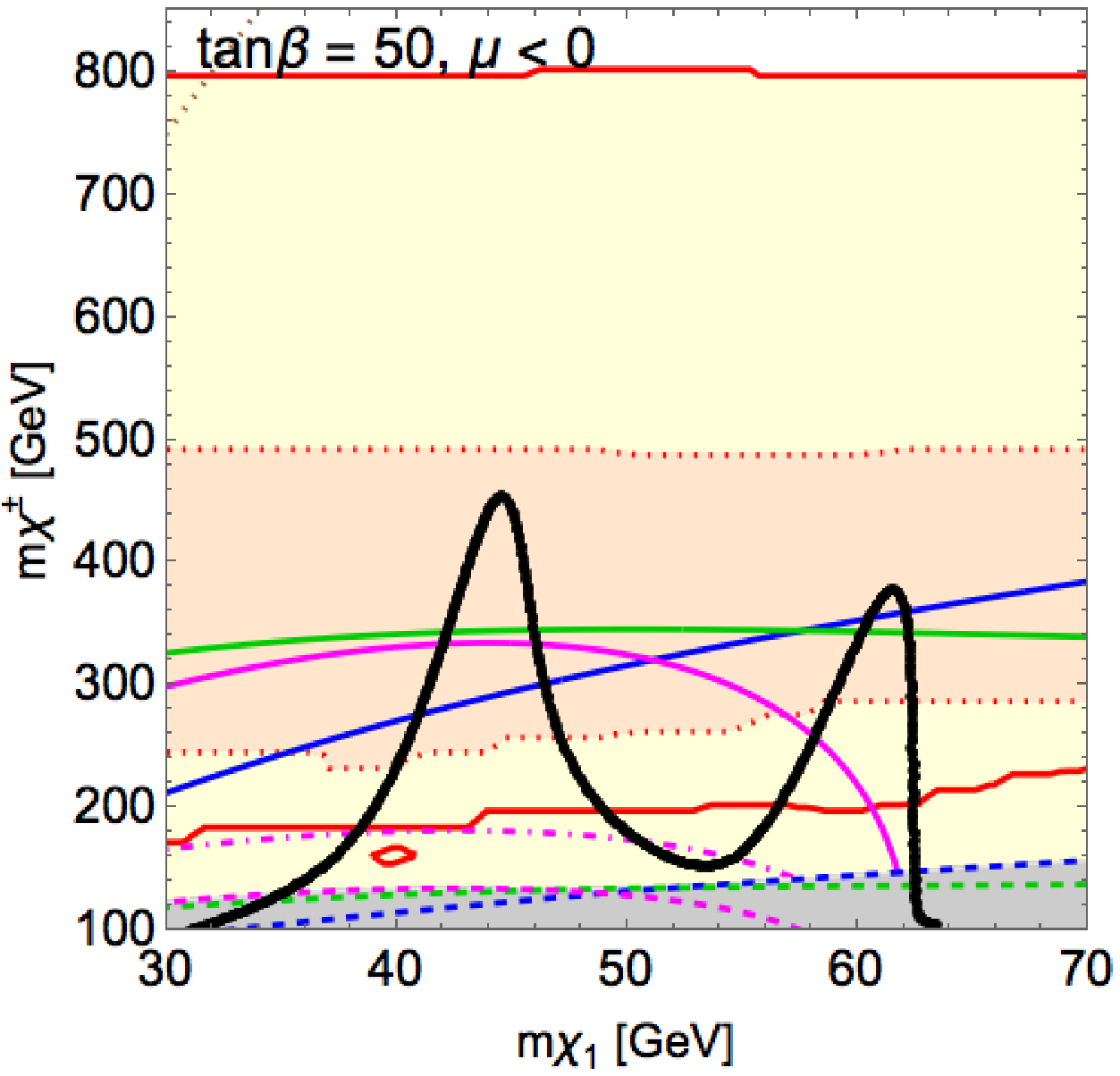}
\includegraphics[width=6.5cm]{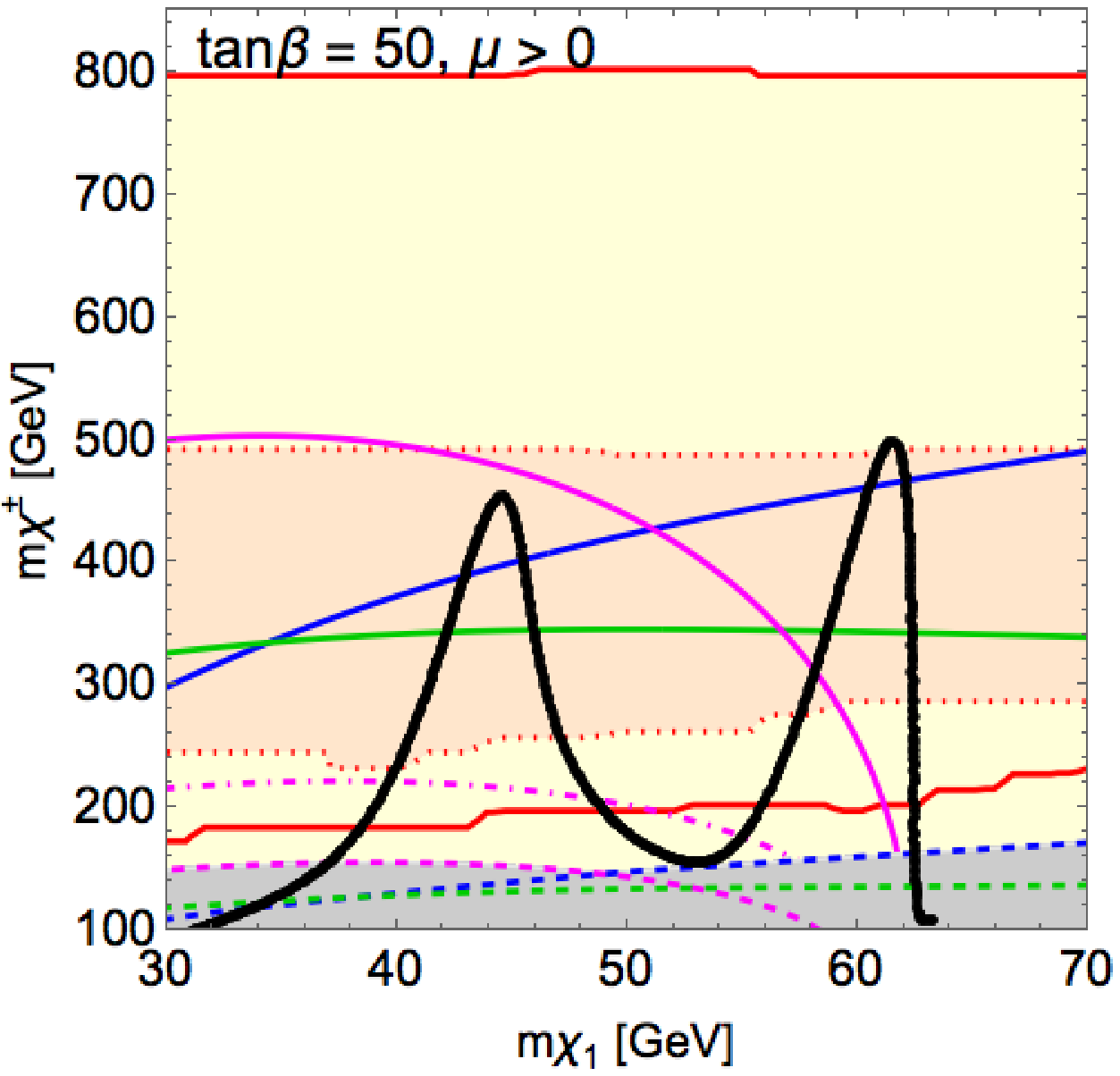}
\caption{The same as Fig.~\ref{fig:main_tb2and3}, but for $\tan\beta=30$, 40 and 50.}
\label{fig:main_tb3040and50}
\end{center}
\end{figure}

The results are summarized in Figs.~\ref{fig:main_tb2and3}--\ref{fig:main_tb3040and50}.
It was shown that there is still a large viable parameter space, and almost all the parameter space of the scenario will be covered complementarily by the LHC search, the direct detection experiments, and the Higgs invisible decay search.

In the $Z$-resonant region, the thermal relic abundance leads to an universal upper bound on the chargino mass, $m_{\chi^\pm}\lesssim 450~\GeV$, independently of the $\text{sign}(\mu)$ and $\tan\beta$.
This region will be covered by the chargino/neutralino searches at the LHC at 300 fb$^{-1}$ except for light chargino region $m_{\chi^\pm}\lesssim 200~\GeV$.
For $\mu>0$, almost all the $Z$-resonant region is probed by both of the XENON 1T and the Higgs invisible decay search at the ILC.
For $\mu<0$, due to the blind spot, there are parameter regions which are not covered by the XENON 1T and/or the Higgs invisible decay search at the ILC, depending on $\tan\beta$ and $m_{\chi_1}$.

In the Higgs resonant region, the upper bound on the chargino mass depends on $\text{sign}(\mu)$ and $\tan\beta$.
For $\mu>0$, larger $\tan\beta$ leads to smaller upper bound on $m_{\chi^\pm}$, e.g., $m_{\chi^\pm}\lesssim 2.5~\TeV$ for $\tan\beta=2$ and $m_{\chi^\pm}\lesssim 500~\GeV$ for $\tan\beta=50$. 
The XENON 1T will cover almost all the region, and the ILC can probe the Higgs invisible decay for the small $m_{\chi^\pm}$ region.
For $\mu<0$, the allowed region has a nontrivial behavior due to the blind spot.
In both of the two cases $\mu>0$ and $\mu<0$, the LHC at 3000 fb$^{-1}$ can cover the region of $m_{\chi^\pm}\lesssim 800~\GeV$.

It is interesting that, depending on $\tan\beta$, $\text{sign}(\mu)$,  $m_{\chi_1}$,  and $m_{\chi_\pm}$, different combinations of positive and negative signals from different experiments may appear.
It is also encouraging that, in the mass range of Higgs- and $Z$-resonant DM, the direct detection experiments may be able to determine the DM mass within certain uncertainty~\cite{1206.6288,Kavanagh:2013wba,Feldstein:2014gza}.

In this paper, we considered a simplified model where all SUSY multiplets except for the Bino and Higgsino multiplets are decoupled and CP is conserved. 
It is interesting to construct a SUSY breaking model to realize such a spectrum and to study the effects of heavier particles and possible CP violation, which are left for future work.

\section*{Acknowledgement}
The authors are thankful to 
Motoi Endo, Jonas Lindert, Takeo Moroi, Mihoko Nojiri, Toshihiko Ota, Satoshi Shirai, Yoshitaro Takaesu, and Takashi Yamanaka for helpful comments and discussions.
K.H. thanks DESY theory group and the Munich Institute for Astro- and Particle Physics (MIAPP), where part of this work was done.
This work was supported by Grant-in-Aid for Scientific research No. 26104001, 26104009, 26247038, 26800123, by World Premier International Research Center Initiative (WPI Initiative), MEXT, Japan, by the MIAPP of the DFG cluster of excellence "Origin and Structure of the Universe", and by the Collaborative Research Center (SFB) 676 at Hamburg University, ``Particles, Strings, and the Early Universe.''

\clearpage

\appendix

\section{Masses, mixings, and couplings}
\label{app:massandmixing}
The masses and mixing angles in Eq.~\eqref{eq:massandmixing} are given by
\begin{align}
m_i &= \frac{1}{3}M_1 + \frac{2}{\sqrt{3}}\mu \sqrt{p}\cos\alpha_i \quad (i=1,2,3)\,,
\\
\OiB{i} &= \frac{ \mu^2-m_{i}^2}{\sqrt{ (\mu^2-m_{i}^2)^2 + (\mu^2 + m_{i}^2 + 4\mu m_{i} c_\beta s_\beta) \widetilde{m}_Z^2}}\,,
\\
\OiHd{i} &= \frac{ \mu s_\beta + m_i c_\beta}{\mu^2-m_{i}^2} \widetilde{m}_Z \OiB{i}\,,
\\
\OiHu{i} &= \frac{- \mu c_\beta -  m_i s_\beta}{\mu^2-m_{i}^2} \widetilde{m}_Z \OiB{i}\,,
\end{align}
where $m_i=\epsilon_i m_{\chi_i}$, $c_\beta=\cos\beta$, $s_\beta=\sin\beta$, $\widetilde{m}_Z=m_Zs_W$, and 
\begin{align}
&\{ \alpha_1,\alpha_2,\alpha_3\}
=
\left\{ \alpha, \alpha+\frac{2\pi}{3}, \alpha+\frac{4\pi}{3} \right\},\quad \alpha = \frac{1}{3}\arccos\left(\frac{\sqrt{27}\;q}{2\;p^{3/2}}\right)\,,
\nonumber\\
&p=1+\frac{1}{3}r_1^2+r_Z^2\,,~
q= -\frac{2}{3}r_1 + \frac{2}{27}r_1^3 + 2 r_Z^2 c_\beta s_\beta+\frac{1}{3} r_1 r_Z^2\,,~
r_1=\frac{M_1}{\mu}\,,~
r_Z=\frac{\widetilde{m}_Z}{\mu}\,.
\end{align}
In terms of ${\cal O}(\widetilde{m}_Z/\mu)$ expansion, the masses and mixings are approximately given by 
\begin{align}
m_1 &= M_1 - \frac{2c_\beta s_\beta + r_1}{1-r_1^2}r_Z^2\; \mu + {\cal O} (r_Z)^3\;\mu 
\\
\OiB{1} &= 1 - \frac{1+r_1^2+4c_\beta s_\beta r_1}{2(1-r_1^2)^2} r_Z^2 + {\cal O} (r_Z)^3\,,
\\
\OiHd{1} &= \frac{s_\beta + c_\beta r_1}{1-r_1^2} r_Z + {\cal O} (r_Z)^3\,,
\\
\OiHu{1} &= \frac{-c_\beta - s_\beta r_1}{1-r_1^2} r_Z + {\cal O} (r_Z)^3\,,
\end{align}
and
\begin{align}
m_j &= \pm \mu \left( 1 + \frac{(c_\beta \pm s_\beta)^2}{2(1\mp r_1)} r_Z^2 + {\cal O} (r_Z)^3\right)\,,
\\
\OiB{j} &= \frac{\mp c_\beta -s_\beta }{\sqrt{2}(1\mp r_1)} r_Z + {\cal O} (r_Z)^3\,,
\\
\OiHd{j} &= \frac{1}{\sqrt{2}} + \frac{2 s_\beta (\mp c_\beta - s_\beta) \mp (c_\beta^2-s_\beta^2)r_1}{4\sqrt{2}(1\mp r_1)^2} r_Z^2 + {\cal O} (r_Z)^3\,,
\label{eq:OuHdj}
\\
\OiHu{j} &= \mp\frac{1}{\sqrt{2}} + \frac{2c_\beta(\pm c_\beta+s_\beta) + (-c_\beta^2+s_\beta^2)r_1}{4\sqrt{2}(1\mp r_1)^2} r_Z^2 + {\cal O} (r_Z)^3\,,
\label{eq:OuHuj}
\end{align}
where $j=2, 3$ depending on the $\text{sign}(\mu)$ and $\tan\beta$.

\section{Constraints from the LHC run I}
\label{app:8TeV}
As discussed in Sec.~\ref{subsec:LHC}, search for the chargino and neutralino at the LHC is sensitive to the present scenario.
At the LHC run I, the process~\eqref{eq:WZatLHC} is searched for by ATLAS~\cite{1402.7029} and CMS~\cite{1405.7570} based on $20.3~\text{fb}^{-1}$ and $19.5~\text{fb}^{-1}$ data at $\sqrt{s}=8~\TeV$, respectively. 
Here, we investigate the constraints from the ATLAS analysis~\cite{1402.7029} in the present scenario.\footnote{Reinterpretation of the ATLAS analysis in the Higgs- and $Z$-resonant neutralino scenario has been done in Ref.~\cite{1410.5730}. See discussion below.}
For this analysis, the CheckMATE program~\cite{1312.2591} is used to evaluate the bounds.\footnote{
CheckMATE uses Delphes 3~\cite{1307.6346}, FastJet~\cite{FastJet}, and the anti-$k_T$ jet algorithm~\cite{Cacciari:2008gp}.
}
As in Sec.~\ref{subsec:LHC}, the events are generated with MadGraph5\_aMC@NLO 2.2.3~\cite{1405.0301} in combination with PYTHIA 6.4~\cite{hep-ph/0603175}.\footnote{We have also checked our results by generating the events with Herwig$++$ 2.7~\cite{Bahr:2008pv}. The results by using MadGraph + PYTHIA  and Herwig++ agree within statistical uncertainties of Monte-Carlo events.}

In the 8 TeV ATLAS analysis~\cite{1402.7029}, many SRs are considered depending on the target model.
Among them, we consider the SR0$\tau$a, which is sensitive to the channel of Eq.~\eqref{eq:WZatLHC} (as well as models with light sleptons).
The SR0$\tau$a is composed of 20 disjoint bins, SR0$\tau$a1--SR0$\tau$a20 defined by different kinematical cuts.
Among them, SR0$\tau$a16 gives the severest constraint in most of the parameter region~\cite{1402.7029}.

First of all, we show the cut flows of some SRs as a validation of our analysis based on the CheckMATE program~\cite{1312.2591}. 
We check the following two model points which are validated by the CheckMATE collaboration~\cite{CheckMATEvalidate};
\begin{align}
(m_{\chi_2},m_{\chi^\pm},m_{\chi_1})=(175,175,100), (350,350,50)~\GeV\,,
\end{align}
in the ``pure Wino'' models.
Here, we assume $\text{Br}(\chi^\pm\to\chi_1 W)= \text{Br}(\chi_j\to\chi_1 Z)=1$ for kinematically allowed region, and $\text{Br}(\chi^\pm\to\chi_1 W^*\to \chi_1\ell\nu)=\text{Br}(W\to \ell\nu)$ and $\text{Br}(\chi_j\to\chi_1 Z^*\to \chi_1\ell\ell)= \text{Br}(Z\to \ell\ell)$ for kinematically forbidden region.

\begin{table}
\begin{center}
\begin{tabular}{c|l|ccc}
\hline \hline
\multicolumn{2}{l|}{Point}
 & \multicolumn{3}{|c}{$m_{\chi_2}=m_{\chi^\pm}=175~\GeV$, $m_{\chi_1}=100~\GeV$}\\
\multicolumn{2}{l|}{Source}
& ATLAS~\cite{1402.7029,HepData}
& CheckMATE~\cite{CheckMATEvalidate}
& our analysis \\
\multicolumn{2}{l|}{Generated events}
& 20000 
& 50000 
& 50000
\\ \hline
\multicolumn{2}{l|}{Initial Events}
& 897 $\pm$ 0 
& 897 $\pm$ 0 
& 897 $\pm$ 0 
\\
\multicolumn{2}{l|}{3 isol. lep., no tau}
& 148 $\pm$ 2.4 
& 142 $\pm$ 1.5 
& 138 $\pm$ 1.6 
\\
\multicolumn{2}{l|}{SFOS, $m_{\text{SFOS}} = 60$--$81.2~\GeV$}
& 78 $\pm$ 1.8 
& 73.9 $\pm$ 1.1 
& 67.7 $\pm$ 1.1 
\\
\multicolumn{2}{l|}{b-veto}
& 75 $\pm$ 1.8 
& 71.1 $\pm$ 1.1 
& 65.8 $\pm$ 1.1 
\\ \hline
SR0$\tau$a9 &
$E^{\text{miss}}_T = 50$--$75~\GeV$ 
& 20 $\pm$ 0.94 
& 19.4 $\pm$ 0.58 
& 17.7 $\pm$ 0.56 
\\
&
$m_T=0$--$80~\GeV$
& 13 $\pm$ 0.76 
& 13.5 $\pm$ 0.49 
& 12.4 $\pm$ 0.47 
\\
&
$|m_{3\ell}-m_Z|>10~\GeV$ 
& 10 $\pm$ 0.67 
& 9.45 $\pm$ 0.41
& 8.16 $\pm$ 0.38 
\\ \hline
SR0$\tau$a10 &
$E^{\text{miss}}_T = 50$--$75~\GeV$
& 20 $\pm$ 0.94 
& 19.4 $\pm$ 0.58 
& 17.7 $\pm$ 0.56 
\\
&
$m_T \geq 80~\GeV$ 
& 7 $\pm$ 0.56 
& 5.95 $\pm$ 0.33 
& 5.47 $\pm$ 0.31 
\\ \hline
SR0$\tau$a11 &
$E^{\text{miss}}_T \geq 75~\GeV$ 
& 19 $\pm$ 0.91 
& 18.8 $\pm$ 0.57 
& 16.7 $\pm$ 0.55
\\
&
$m_T = 0$--$100~\GeV$ 
& 15 $\pm$ 0.81 
& 15.7 $\pm$ 0.53 
& 13.9 $\pm$ 0.50 
\\ \hline
SR0$\tau$a12 &
$E^{\text{miss}}_T \geq 75~\GeV$ 
& 19 $\pm$ 0.91 
& 18.8 $\pm$ 0.57
& 16.7 $\pm$ 0.55 
\\
&
$m_T \geq 100~\GeV$ 
& 4 $\pm$ 0.42 
& 3.07 $\pm$ 0.23 
& 2.93 $\pm$ 0.23
\\ \hline
 \hline
 \multicolumn{4}{c}{}\\
 \hline \hline
\multicolumn{2}{l|}{Point} & \multicolumn{3}{|c}{$m_{\chi_2}=m_{\chi^\pm}=350~\GeV$, $m_{\chi_1}=50~\GeV$}\\
\multicolumn{2}{l|}{Source}
& ATLAS~\cite{1402.7029,HepData}
& CheckMATE~\cite{CheckMATEvalidate}
& our analysis  \\
\multicolumn{2}{l|}{Generated events}
& 20000 
& 50000 
& 50000\\ \hline
\multicolumn{2}{l|}{Initial Events}
& 49.2 $\pm$ 0 
& 49.2 $\pm$ 0 
& 49.2 $\pm$ 0 \\
\multicolumn{2}{l|}{3 isol. lep., no tau}
& 11 $\pm$ 0.14 
& 11.9 $\pm$ 0.094 
& 11.7 $\pm$ 0.11\\
\multicolumn{2}{l|}{SFOS, $m_{\text{SFOS}} = 81.2$--$101.2~\GeV$}
& 10 $\pm$ 0.14 
& 9.87 $\pm$ 0.088 
& 9.47 $\pm$ 0.097\\
\multicolumn{2}{l|}{b-veto}
& 10 $\pm$ 0.14 
& 9.39 $\pm$ 0.086 
& 9.11 $\pm$ 0.095\\ \hline
SR0$\tau$a13 &
$E^{\text{miss}}_T = 50$--$90~\GeV$ 
& 1.1 $\pm$ 0.051 
& 1.03 $\pm$ 0.031 
& 1.03 $\pm$ 0.032\\
&
$m_T=0$--$110~\GeV$ 
& 0.6 $\pm$ 0.038 
& 0.673 $\pm$ 0.026 
& 0.671 $\pm$ 0.026\\
&
$|m_{3\ell}-m_Z|>10~\GeV$ 
& 0.6 $\pm$ 0.038 
& 0.665 $\pm$ 0.025 
& 0.665 $\pm$ 0.026\\ \hline
SR0$\tau$a14 &
$E^{\text{miss}}_T \geq 90~\GeV$ 
& 8 $\pm$ 0.13 
& 7.87 $\pm$ 0.081 
& 7.63 $\pm$ 0.0866\\
&
$m_T = 0$--$110~\GeV$ 
& 2.4 $\pm$ 0.075 
& 2.53 $\pm$ 0.049
& 2.34 $\pm$ 0.048\\ \hline
SR0$\tau$a15 &
$E^{\text{miss}}_T = 50$--$135~\GeV$ 
& 2.9 $\pm$ 0.082 
& 2.78 $\pm$ 0.051
& 2.66 $\pm$ 0.051\\
&
$m_T \geq 110~\GeV$ 
& 1.6 $\pm$ 0.062 
& 1.29 $\pm$ 0.035 
& 1.25 $\pm$ 0.035\\ \hline
SR0$\tau$a16 &
$E^{\text{miss}}_T \geq 135~\GeV$ 
& 7 $\pm$ 0.12 
& 6.12 $\pm$ 0.073 
& 6.00 $\pm$ 0.077\\
&
$m_T \geq 110~\GeV$ 
& 5 $\pm$ 0.11 
& 4.4 $\pm$ 0.063
& 4.40 $\pm$ 0.066\\ \hline
 \hline
\end{tabular}
\caption{Cut-flow validation.
The errors are statistics of Monte-Carlo events only. }
\label{table:cutflow}
\end{center}
\end{table}

In Table.~\ref{table:cutflow}, we show the cut-flow validation for these model points, and compare them with the CheckMATE validation~\cite{CheckMATEvalidate} and the ATLAS cut-flow~\cite{1402.7029,HepData}. 
The initial event number is normalized to the one of ~\cite{CheckMATEvalidate}.
At first, leptons, jets and missing transverse energy are defined as in~\cite{1402.7029}.
One of the triggers in ~\cite{1402.7029} should be satisfied.
Then, the following cuts are applied.
\begin{itemize}
\item Exactly three isolated leptons with no taus are required.
\item At least one pair of same flavor opposite sign (SFOS) leptons should exist. 
Among the SOFS pairs, the SOFS mass which is closest to the $Z$-boson mass should be in the range defined in each SR, e.g. $m_{\text{SFOS}}=60$--$81.2~\GeV$ for SR0$\tau$a9--12 and $m_{\text{SFOS}}=81.2$--$101.2~\GeV$ for SR0$\tau$a13--16.
\item Events including the b-tagged jets are vetoed.
\item The events are further divided into four bins depending on the missing transverse energy $E_T^{\text{miss}}$ and the transverse mass $m_T$ (see Table.~\ref{table:cutflow}), where $m_T$ is calculated with missing energy and the lepton which does not form the SFOS lepton pair whose mass is closest to the $Z$-boson mass.
\item In some SRs, additional requirement on the trilepton mass, $|m_{3\ell}-M_Z| > 10~\GeV$, is applied. 
\end{itemize}
As can be seen in Table.~\ref{table:cutflow}, although in some SRs the intermediate cut-flows are different, the overall acceptances agree well with the ones of~\cite{CheckMATEvalidate}.
In particular, in the SR0$\tau$a16, which gives the severest constraint in most of the parameter region~\cite{1402.7029}, the acceptance in our analysis agrees very well with the one in~\cite{CheckMATEvalidate}.

\begin{table}
\begin{center}
\begin{tabular}{|l|cc|cc|}
\hline
$m_{\chi^\pm},m_{\chi_1}$ [GeV]
& $\sigma^{\text{NLO}}_{\chi^\pm \chi_j}$ [fb]
& $\sigma^\text{ATLAS}_{\chi^\pm \chi_j}$ [fb]
& $A_{\text{SR0}\tau\text{a16}}\times 10^3$
& $A_{\text{SR0}\tau\text{a16}}^\text{ATLAS}\times 10^3$
\\ \hline
200, 50 & 788 & 802 & 15.5 $\pm$ 0.2 & 18.5 \\
200, 25 & 788 & 802 & 20.3 $\pm$ 0.1 & 24.0 \\
150, 37.5 & 2427 & 2452 & 2.26 $\pm$ 0.07 & 2.71 \\ \hline
\end{tabular}
\caption{Comparison of the cross section and the acceptance of our analysis with those of the ATLAS analysis~\cite{HepData} for the ``pure Wino'' models.
}
\label{table:vsATLAS}
\end{center}
\end{table}

Secondly, in Table~\ref{table:vsATLAS}, we compare the cross section and the acceptance of our analysis with those of the ATLAS analysis for the ``pure Wino'' model points $(m_{\chi^\pm},m_{\chi_1})=(200,50), (200,25)$ and $(150,37.5)~\GeV$.\footnote{
These model points are chosen because the efficiency, the acceptance, and the production cross section for the ``pure Wino" case in the ATLAS analysis are available at ~\cite{HepData}.
}
Here, the acceptance $A_{\text{SR0}\tau\text{a16}}$ is defined as in Eq.~\eqref{eq:acc2}, and the effective acceptance of the ATLAS corresponding to Eq.~\eqref{eq:acc2} is calculated as $A_{\text{SR0}\tau\text{a16}}^\text{ATLAS}=[A\cdot \epsilon]_{\text{SR0}\tau\text{a16}}^\text{ATLAS}\cdot \text{Br}(Z\to \ell\ell)^{-1}\text{Br}(W\to \ell\nu)^{-1}$, where the acceptance times the efficiency $[A\cdot \epsilon]_{\text{SR0}\tau\text{a16}}^\text{ATLAS}$ is taken from HepData~\cite{HepData}.\footnote{In our simulation, the contributions from the hadronic decays of $W$ and $Z$ are negligibly small.
}
Compared to the ATLAS analysis~\cite{1402.7029,HepData}, the estimated acceptance is about 20\% smaller.
The cross sections are well reproduced within 1--2\%.

\begin{figure}[t]
\begin{center}
\includegraphics[]{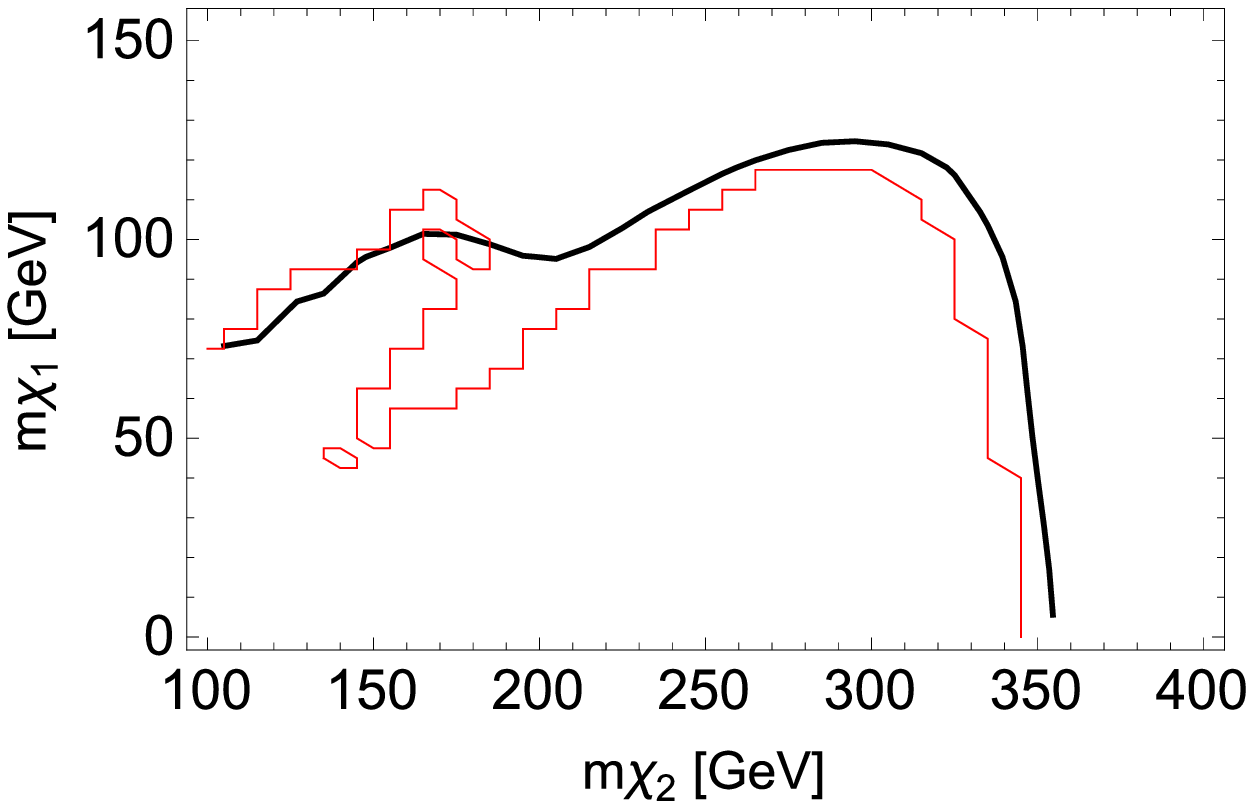}
\caption{Reinterpretation of the ATLAS analysis ~\cite{1402.7029} using the CheckMATE program~\cite{1312.2591}.
The black line shows the ATLAS result given in ~\cite{HepData}.
The red line denotes the result of our analysis .
}
\label{fig:validation}
\end{center}
\end{figure}

We have performed the same analysis in the $(m_{\chi_2},m_{\chi_1})$-plane, and show the exclusion contour in Fig.~\ref{fig:validation}.
Here, all the 20 SRs (SR0$\tau$a1--20) are taken into account.
The ATLAS result~\cite{1402.7029} is shown in black line.
The red line denotes the result of our analysis.
Although the shape near the kinematical edge is slightly different, 
the result of our analysis is in good agreement with the ATLAS analysis.

\begin{table}
\hspace{-1.5cm}
\begin{tabular}{|c|cc|cccc|c|c|}
\hline
$m_{\chi^\pm},m_{\chi_1}$ & $\tan\beta,\text{sign}(\mu)$ 
& $\chi_j$
& $m_{\chi_j}$
& $\sigma(\chi^\pm \chi_j)$
& $\text{Br}(WZ)$ 
& $A_{\text{SR0}\tau\text{a16}}\times10^3$
&  $N_{\text{SR0}\tau\text{a16}}$
& $N_{\text{SR0}\tau\text{a16}}^{\text{All}}$
\\ \hline\hline
& 5, $+$ & $\chi_2$  & 202.3  & 192  & 0.958  & 16.9 $\pm$ 0.2 & 2.08 $\pm$ 0.02  &
\\& & $\chi_3$  & 208.8  & 171  & 0.328  & 18.7 $\pm$ 0.2  & 0.70  $\pm$ 0.01 & 2.91 $\pm$ 0.10 
\\ \cline{2-9}
& 5, $-$ & $\chi_2$  & 203.9 & 185 &  0.697 & 18.0 $\pm$ 0.2  &  1.55 $\pm$ 0.02 &
\\& & $\chi_3$  & 205.3 &  183  &  0.819 & 18.2 $\pm$ 0.2  & 1.82 $\pm$ 0.02 & 3.75 $\pm$ 0.12 
\\ \cline{2-9}
200, 50 & 40,  $+$ & $\chi_2$  & 203.6  & 188 & 0.919 & 17.7 $\pm$ 0.2  & 2.05 $\pm$ 0.02 &
\\& & $\chi_3$  & 206.7 & 177 &  0.490 & 18.1 $\pm$ 0.2  & 1.05 $\pm$ 0.01 & 3.25 $\pm$ 0.11
\\ \cline{2-9}
& 40, $-$ & $\chi_2$ & 204.0 & 186 &  0.902 & 18.0 $\pm$ 0.2  & 2.01 $\pm$ 0.02 &
\\& & $\chi_3$ & 206.0 &  179 &  0.538 & 18.5 $\pm$ 0.2 &1.19 $\pm$ 0.01 & 3.53 $\pm$ 0.12
\\ \cline{2-9}
& ``pure Wino" & $\chi_2$  & 200  & 788 &   1.0 & 15.5 $\pm$ 0.2  & 8.15 $\pm$ 0.09 &
\\ \hline
%%%%%%%%%%%%%%%%%%%%%%%%%%%%%%%%%
& 5, $+$ & $\chi_2$  & 202.6 & 191 & 0.902 & 22.4$\pm$ 0.2  & 2.58 $\pm$ 0.02 &
\\& & $\chi_3$  & 207.6 & 176  &  0.405  &  24.1$\pm$ 0.2 & 1.14 $\pm$ 0.01 & 3.84 $\pm$ 0.12 
\\ \cline{2-9}
& 5, $-$ & $\chi_2$  & 203.3 & 188 &  0.761  & 23.3 $\pm$ 0.2  & 2.23 $\pm$ 0.02 &
\\& & $\chi_3$  & 205.9 & 182 &  0.657 & 23.7  $\pm$ 0.2  & 1.89 $\pm$ 0.02 & 4.49 $\pm$  0.13
\\ \cline{2-9}
200, 25 & 40,  $+$ & $\chi_2$  & 204.0 & 186 &  0.822  &  23.1 $\pm$ 0.2  & 2.35 $\pm$ 0.02 &
\\& & $\chi_3$  & 205.7 &  180 &  0.572  &  23.4 $\pm$ 0.2  & 1.61 $\pm$ 0.01 & 4.18 $\pm$ 0.13
\\ \cline{2-9}
& 40, $-$ & $\chi_2$  & 204.4 & 186  &  0.792  &  22.8 $\pm$ 0.2  & 2.24 $\pm$ 0.02 &
\\& & $\chi_3$  & 205.2 & 182 & 0.618  &  23.1 $\pm$ 0.2  & 1.74 $\pm$ 0.02 & 4.47 $\pm$ 0.13
\\ \cline{2-9}
& ``pure Wino" & $\chi_2$  & 200  & 788  &   1.0 &  20.3 $\pm$ 0.1  & 10.7 $\pm$ 0.1 &
\\ \hline
%%%%%%%%%%%%%%%%%%%%%%%%%%%%%%%%%
& 5, $+$ & $\chi_2$  & 153.0 &  575 & 1.0   & 2.76 $\pm$ 0.07  & 1.06 $\pm$ 0.03 &
\\& & $\chi_3$  & 161.7 &  474 &  1.0   & 3.67 $\pm$ 0.09  &  1.16 $\pm$ 0.03 & 2.41 $\pm$ 0.16
\\ \cline{2-9}
& 5, $-$ & $\chi_2$  & 155.1 &  545 &   1.0  & 3.06 $\pm$ 0.08 & 1.11 $\pm$ 0.03 &
\\& & $\chi_3$  & 157.0 &  536 &   1.0  & 3.15 $\pm$ 0.08  & 1.13 $\pm$ 0.03 & 2.46 $\pm$ 0.17
\\ \cline{2-9}
150, 37.5 & 40,  $+$ & $\chi_2$  & 154.7 &   558 & 1.0 & 3.08 $\pm$ 0.08  & 1.15 $\pm$ 0.03 &
\\& & $\chi_3$  & 158.8 &  503 &  1.0 & 3.42 $\pm$ 0.08  & 1.15 $\pm$ 0.03 & 2.68 $\pm$ 0.17
\\ \cline{2-9}
& 40, $-$ & $\chi_2$  & 155.2 & 553 &  1.0& 3.23 $\pm$ 0.08  & 1.19 $\pm$ 0.03 &
\\& & $\chi_3$  & 158.0 & 513 &  1.0& 3.49 $\pm$ 0.08  & 1.19 $\pm$ 0.03 & 2.43 $\pm$ 0.16
\\ \cline{2-9}
& ``pure Wino" & $\chi_2$  & 150  & 2427 &   1.0 & 2.26 $\pm$ 0.07  & 3.66 $\pm$ 0.11 &
\\ \hline
%%%%%%%%%%%
\end{tabular}
\caption{
The masses of heavier neutralinos $m_{\chi_{2,3}}$, the NLO production cross sections $\sigma(\chi^\pm \chi_j)=\sum_{\chi^\pm}\sigma^{\text{NLO}}(pp\to \chi^\pm \chi_j)$, the branching ratio of the $WZ$ mode
$\text{Br}(WZ)=\text{Br}(\chi_j\to \chi_1Z)$, 
the acceptance $A_{\text{SR0}\tau\text{a16}}$ of SR0$\tau$a16, and the expected number of signal events $N_{\text{SR0}\tau\text{a16}}$ in  SR0$\tau$a16, for the model points of Eq.~\eqref{eq:12points}. 
For comparison, we also show the results for ``pure Wino" models.
$A_{\text{SR0}\tau\text{a16}}$ are calculated as in Eq.~\eqref{eq:acc2}, and their errors are statistics of Monte-Carlo events only.
$N_{\text{SR0}\tau\text{a16}}$ is calculated by using Eq.~\eqref{eq:NSRX}.
$N_{\text{SR0}\tau\text{a16}}^{\text{All}}$ is the expected number including all the production and decay channels.
The masses and the cross sections are in units of [GeV] and [fb], respectively.
}
\label{table:all}
\end{table}

Next, let us reinterpret this analysis to the present scenario.
In Table.~\ref{table:all}, we show (i) the masses of heavier neutralinos $m_{\chi_{2,3}}$, 
(ii) the NLO production cross sections $\sum_{\chi^\pm}\sigma^{\text{NLO}}(pp\to \chi^\pm \chi_j)$, 
(iii) the branching ratio of the $WZ$ mode $\text{Br}(WZ)=\text{Br}(\chi_j\to \chi_1Z)$\footnote{
Note that $\text{Br}(\chi^\pm\to \chi_1W)=1.0$.},
(iv) the acceptance $A_{\text{SR0}\tau\text{a16}}$ of SR0$\tau$a16, and (v) the expected number of signal events $N_{\text{SR0}\tau\text{a16}}$ in SR0$\tau$a16, for the following 12 model points in the present scenario.
\begin{align}
&(m_{\chi^\pm},m_{\chi_1})=(200,50), (200,25), (150,37.5)~\GeV\,,
\nonumber\\
&\tan\beta = 5, 40\,,\qquad \text{sign}(\mu)=\pm\,.
\label{eq:12points}
\end{align}
The masses,  mixing angles, and branching ratios are calculated at tree level.
For comparison, we also show the case of ``pure Wino'' models with $\text{Br}(\chi_2\to \chi_1Z)=1$.
As shown in the Table~\ref{table:all}, the acceptance in the present scenario is slightly better than the ``pure Wino" case, and larger $m_{\chi_{2,3}}$ lead to larger acceptance.
However, the production cross section, $\sum_{j=2,3}\sigma^\text{NLO}(\chi^\pm\chi_j)$, is about one half of the ``pure Wino" case.
Furthermore, the branching fraction of the $WZ$ mode is smaller for $(m_{\chi^\pm},m_{\chi_1})=(200,50)$ and (200,25)~GeV.
As a result, the expected number of signal events in SR0$\tau$a16, $N_{\text{SR0}\tau\text{a16}}$, becomes less than about 40\% and 65\% of the ``pure Wino" case for $(m_{\chi^\pm},m_{\chi_1})=(200,50/25)$ GeV and (150, 37.5)~GeV, respectively.

So far, we have considered only the production and decay modes in Eq.~\eqref{eq:WZatLHC}.
In order to check the contributions from the other channels, we have generated all the possible pair production channels, $pp\to \chi_i\chi_j, \chi_i \chi^\pm, \chi^+\chi^-$, ($i, j=1,2,3$) and also included the decay into the Higgs boson, $\chi_{2,3} \to h\chi_{1}$, as well as the hadronic decays of $W$ and $Z$. 
The resultant overall expected number of signal events in SR0$\tau$a16, $N_{\text{SR0}\tau\text{a16}}^{\text{All}}$, are shown in the last column of Table~\ref{table:all}.\footnote{
We have generated 3,000,000 events for each model point to calculate $N_{\text{SR0}\tau\text{a16}}^{\text{All}}$.
}
They are at most about 15\% larger than $N_{\text{SR0}\tau\text{a16}}$.
We have checked that the additional contributions mainly come from the production channel $pp\to \chi_2\chi_3$.

\begin{table}
\begin{center}
\begin{tabular}{|c|cccc|cccc|cccc|c|}
\hline
$(m_{\chi^\pm},m_{\chi_1})$
&\multicolumn{4}{|c|}{(200, 50)  [GeV]}
&\multicolumn{4}{|c|}{(200, 25)  [GeV]}
&\multicolumn{4}{|c|}{(150, 37.5)  [GeV]}
&\\
$\tan\beta$
& 5 & 5 & 40 & 40
& 5 & 5 & 40 & 40
& 5 & 5 & 40 & 40
&\\
sign($\mu$)
& $+$ & $-$ & $+$ & $-$
& $+$ & $-$ & $+$ & $-$
& $+$ & $-$ & $+$ & $-$
& $N_{\text{obs}}^{95}~\cite{1402.7029}$
\\\hline 
$N_{\text{SR0}\tau\text{a14}}^{\text{All}}$
&7.2 &7.7 &7.9 &8.1
&7.7 &8.2 &8.2 &8.1
&21.9 &22.9 &22.2 &22.4
& 65 \\\hline 
$N_{\text{SR0}\tau\text{a15}}^{\text{All}}$
&6.9 &8.2 &7.5 &7.9
&7.2 &7.8 &7.4 &8.0
&19.2 &21.0 &21.2 &21.0
& 27.6 \\\hline 
$N_{\text{SR0}\tau\text{a16}}^{\text{All}}$
&2.9 &3.8 &3.3 &3.5
&3.8 &4.5 &4.2 &4.5
&2.4 &2.5 &2.7 &2.4
& 5.2 \\\hline
\end{tabular}
\caption{
The expected number of signal events in SR0$\tau$a14, 15, and 16 of the ATLAS analysis~\cite{1402.7029} for the model points of Eq.~\eqref{eq:12points}. 
$N_{\text{obs}}^{95}$ is the observed upper limits at 95\% CL on the number of beyond-the-SM events for each signal region~\cite{1402.7029}.
}
\label{table:number}
\end{center}
\end{table}

In Table.~\ref{table:number}, we show the expected number of signal events in SR0$\tau$a14, 15, and 16.
Here, we have included all the production and decay channels discussed above.
We compare them with the observed upper limits at 95\% CL on the number of beyond-the-SM events for each signal region, $N_{\text{obs}}^{95}$~\cite{1402.7029}.
In the other signal regions, SR0$\tau$a1--13 and 17--20, the signal events are less than about 10\% and 25\% of the upper limits for $(m_{\chi^\pm},m_{\chi_1})=(200,50/25)$ GeV and (150, 37.5)~GeV, respectively.
We find that none of these model points are excluded.

We have performed the same analysis in the parameter space of Figs.~\ref{fig:main_tb2and3}--\ref{fig:main_tb3040and50},\footnote{
The cross sections are calculated at $\tan\beta=2,5,50$, sgn($\mu)=\pm$, and $M_1=30, 40,\cdots 80~\GeV$ for $|\mu|=100, 110, \cdots 200~\GeV$, and at $(\tan\beta,\text{sign}(\mu),M_1)=(2,+,80~\GeV), (50,+,30~\GeV)$  for $|\mu|=200, 210, \cdots 400~\GeV$.
Then, the cross sections normalized by the coupling, $\sigma^{\text{NLO}}/(|\OiHu{j}|^2+|\OiHd{j}|^2)$, are interpolated.
The acceptances are calculated by varying the masses by $(\Delta m_{\chi^\pm}, \Delta (m_{\chi_j}-m_{\chi^\pm}), \Delta m_{\chi_1})=(10,10,5)~\GeV$, while the couplings are fixed as the ones of $\tan\beta=5, M_1=50~\GeV, \mu=200~\GeV$, for simplicity. 
(We have checked that the acceptance does not depend much on these parameters.)
}
and found that the parameter regions allowed by the other constraints are not excluded by the 8 TeV LHC constraints.
This result does not agree with the previous work~\cite{1410.5730} where $m_{\chi^\pm} \lesssim 250$ GeV is excluded depending on $\tan\beta$ and $m_{\chi_1}$.\footnote{
We have also analyzed the pMSSM scenario studied in the ATLAS~\cite{1402.7029}, and obtained a weaker constraint.}
We should emphasize that the expected number of signal events and the observed upper limits are the same order in a large region of the parameter space, and hence ${\cal O}(10\%)$ change of the event numbers would drastically change the bounds on the parameter space.
We have checked that our analysis would lead to similar bounds as in~\cite{1410.5730} if the event numbers are increased by about 50\%.
A large part of the parameter region in Figs.~\ref{fig:main_tb2and3}--\ref{fig:main_tb3040and50} is in any case still viable, and will be probed in future experiments as discussed in this paper.

\end{document}